\documentclass[letterpaper,twocolumn,10pt]{article}

\usepackage{usenix,epsfig,endnotes}
\usepackage{amssymb,amsmath,amsthm}
\newtheorem{theorem}{Theorem}
\usepackage{algorithm}
\usepackage{algpseudocode}
\usepackage{subfig}
\usepackage{url}
\usepackage{color}

\begin{document}

\date{}

\title{%
\Large \bf \underline{S}tart \underline{L}ate or \underline{F}inish \underline{E}arly:
A  Distributed Graph Processing System with Redundancy Reduction%
}

\author{
{
\rm $^\dagger$Shuang Song$^1$,
$^\dagger$Xu Liu$^2$
\thanks{$^\dagger$ Shuang Song and Xu Liu are the corresponding authors.},
\ Qinzhe Wu$^1$,
Andreas Gerstlauer$^1$,
Tao Li$^3$,
and Lizy K. John$^1$
}\\
University of Texas at Austin$^1$ 
\\College of William and Mary$^2$
\\University of Florida$^3$
} % end author

\maketitle

\thispagestyle{empty}

\subsection*{Abstract}
Graph processing systems are important in the big data domain.
However, processing graphs in parallel often introduces redundant computations in existing algorithms and models.
Prior work has proposed techniques to optimize redundancies for the out-of-core graph systems,
rather than the distributed graph systems.
In this paper,
we study various state-of-the-art distributed graph systems
and observe root causes for these pervasively existing redundancies.
To reduce redundancies without sacrificing parallelism, 
we further propose 
\textit{SLFE}, %(pronounced as ``Selfie'') 
a distributed graph processing system,
designed with the principle of ``start late or finish early".
\textit{SLFE} employs a novel preprocessing stage to obtain a graph's topological knowledge with negligible overhead.
\textit{SLFE}'s redundancy-aware vertex-centric computation model
can then utilize such knowledge to reduce the redundant computations at runtime.
\textit{SLFE} also provides a set of APIs to improve the programmability.
Our experiments on an 8-node high-performance cluster show
that \textit{SLFE} outperforms all well-known distributed graph processing systems on real-world graphs
(yielding up to 74.8$\times$ speedup).
\textit{SLFE}'s redundancy-reduction schemes are generally applicable to other vertex-centric graph processing systems.

\section{Introduction}
\label{sec:intro}
The amount of data generated every day is growing exponentially in the big data era.
By 2020, the digital data volume stored in the world is expected to reach 44 zettabytes~\cite{shuangicpp,baru2013setting}.
A significant portion of this data is stored as graphs in various domains,
such as online retail, social networks, and bio-informatics~\cite{ammar2014wgb}.
Hence, developing distributed systems to efficiently analyze large-scale graphs that cannot fit in a single commodity PC
has been a popular topic in the recent years. 

To achieve high performance,
existing graph systems aim to exploit massive parallelism using either
distributed~\cite{pregel,graphlab,powergraph,geminigraph,gps,Cyclops,chaos,graphx,galois,powerswitch}
or shared memory models~\cite{graphchi,xstream,Nguyen,ligra}.
Such graph systems process graphs in a repeated-relaxing manner
(e.g., using Bellman-Ford algorithm variants~\cite{bellmanford})
rather than in a sequential but work-optimal order.
This introduces a fundamental trade-off between available parallelism and redundant computations~\cite{dsmr1,dsmr2}. 
We study several popular graph processing systems with applications implemented atop them~\cite{geminigraph,powergraph,powerlyra},
and find that redundant computations pervasively exist.
We elaborate on this in Section~\ref{sec:gp}.

According to our investigation,
the root causes of computational redundancies existing in graph analytics vary across applications,
which is caused by the nature
(i.e., core aggregation function)
of different graph algorithms.
As an example, applications such as Single Source Shortest Path (SSSP)
employ $min()$/$max()$ function as their core aggregation function.
In each iteration,
the values of active neighboring vertices are fed into the $min()$/$max()$ aggregation function,
and the result is assigned to the destination vertex.
Typically, a vertex needs multiple value updates in different iterations,
because the value updates in any source vertices require to recompute the destination vertex's property.
However, only one minimum or maximum value is needed to maintain the algorithm's functionality.
Therefore, we define that the redundancies in these applications are the computations triggered by the updates with intermediate (not final min/max) values.
The philosophy of ``start late'' we propose can bypass such redundant updates,
yielding speedups for this type of graph applications.

By contrast, some other graph applications (e.g., PageRank (PR)) utilize the arithmetic operations (e.g., $sum()$) to accumulate the neighbors' values iteratively until no vertex has further changes (a.k.a final convergence). 
For this kind of algorithms,
there are no computational redundancies caused by intermediate updates.
However, our analysis shows that most vertices are early converged
(the vertex's value is stabilized)
before a graph's final convergence.
Hence, following computations on such early-converged (EC) vertices are considered as redundancies for this type of applications.
Because the redundancies occur after the vertex becomes stabilized,
we propose a ``finish early'' approach on EC vertices to eliminate such redundancies.

We develop \textit{SLFE} (pronounced as ``Selfie''),
a distributed graph processing system that reduces redundancies to achieve high performance of different types of graph analytics. 
To reduce the redundancy, we design a novel preprocessing phase
that can produce a graph's topological guidance to guide vertex-centric operations in the following execution phase. 
Such guidance can be utilized by applications using either min()/max() operation or arithmetic operation as aggregation functions.
Compared to the prior work
that leverages dynamic re-sharding/partitioning~\cite{rajiv1} or multi-round partitioning~\cite{rajivhpdc} for redundancy reductions in out-of-core graph systems,
our strategy has the following benefits:
1) it does not incur any extra partitioning effort~\footnote{The partitioning phase in distributed graph systems is expensive~\cite{vermavldb17,stantonkdd12,luvldb14,geminigraph}.}; 
2) it does not rely on any specific ingress methodology, so it can be easily adopted by other systems;
3) it has extremely low preprocessing overhead, which is suitable for on-line optimization;
and 4) it produces guidance that is reusable by various graph algorithms for the redundancy optimizations.

To balance the communication and computation on the fly, \textit{SLFE} uses the state-of-the-art ``push/pull'' computation model. The ``push'' sends the update of source vertices to their successors, while ``pull'' extracts information from predecessors for a given destination vertex.
To the best of our knowledge, \textit{SLFE} is the first graph system with a set of redundancy-reduction aware ``push/pull'' functions to make use of guidance produced in preprocessig. 
Moreover, \textit{SLFE} also provides a set of system APIs to enable redundancy reductions as well as programming simplicity/flexibility for different graph applications.
We summarize the contributions of this paper as follows:

\begin{itemize}
\item We perform a thorough study on state-of-the-art graph processing systems and observe the pervasive existence of large amounts of computational redundancies. We further identify the provenance of these redundancies.  
\item We design a novel and lightweight preprocessing technique to extract a graph's topological information (i.e., propagation order). This technique enables both ``start late'' and ``finish early'' redundancy reduction principles for many graph applications.
\item We develop \textit{SLFE}, a distributed graph processing system that employs various techniques to demonstrate the benefit from optimizing redundancies in graph applications.
\item We evaluate \textit{SLFE} with extensive experiments and compare it with three state-of-the-art distributed graph processing systems. Experiments with five popular applications on seven real-world graphs show that \textit{SLFE} significantly outperforms these systems, yielding speedups up to 74.8$\times$ (16.5$\times$ on average). 
\end{itemize}

\section{Observations and Motivation}
\label{sec:gp}
\begin{table}[t]
\renewcommand{\arraystretch}{1.3}
\caption{A list of graph analytical applications with two different aggregation functions~\cite{rajiv1}.}
\label{table:apps}
\centering
\scriptsize
\begin{tabular}{|c|c|}
\hline
Graph Analytical App & Aggregation Function\\
\hline
PageRank, NumPaths, SpMV,&\\
TriangleCounting, BeliefPropagation, & Arithmetic ({\it sum} or {\it product})\\
HeatSimulation, TunkRank&\\
\hline
SingleSourceSP, MinimalSpanningTree,&\\
ConnectedComponents, WidestPath,& Comparsion ({\it min} or {\it max})\\
ApproximateDiameter, Clique&\\
\hline
\end{tabular}
\vspace{-0.2cm}
\end{table}

\subsection{Graph Applications}
Most popular graph applications can be classified into two categories based on the type of their aggregation functions.
The core computations in aggregation functions of such applications are either arithmetic operation or $min$/$max$ comparison.
We analyze graph applications implemented atop several graph processing systems~\cite{graphchi,graphlab,xstream,ligra,graphx,powergraph,powerlyra,pregel} and summarize our findings in Table~\ref{table:apps}. 
\textit{SLFE} aims to provide a unified solution to reduce the computational redundancies for both types of applications.
We select SSSP and PR to discuss the various provenance of computational redundancies and motivate the \textit{SLFE} design.

\subsection{Computational Redundancy}
We further look into the implementation of graph applications and processing systems. We observe that state-of-the-art graph systems prefer to execute graph applications in a Bellman-Ford~\cite{bellmanford} way to utilize the massive parallelism provided by the underlying computation units. Such implementations often introduce computational redundancies for graph applications with heavyweight $min$/$max$ or arithmetic operations. 

\begin{figure}[t]
\begin{minipage}{0.45\linewidth}
\centering
 \subfloat[An example graph]{\label{fig:example_sssp}\includegraphics[height=0.14\textheight,width=1\linewidth,trim = 0mm 85mm 230mm 0mm, clip=true]{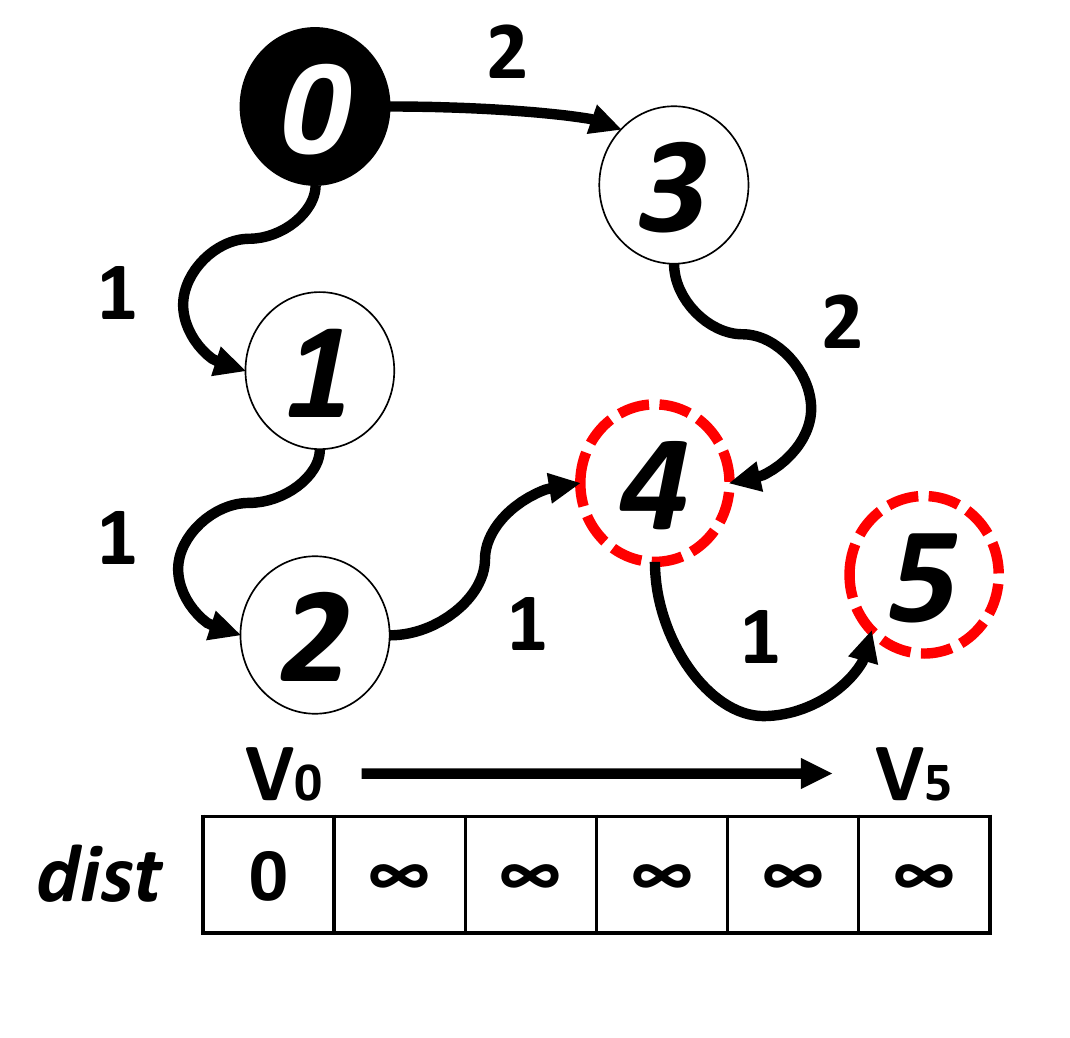}}
\end{minipage}%
\begin{minipage}{0.55\linewidth}
\centering
\subfloat[SSSP's iteration plot]{\label{fig:sssp_iteration}\includegraphics[height=0.14\textheight,width=1\linewidth,trim = 2mm 80mm 200mm 0mm, clip=true]{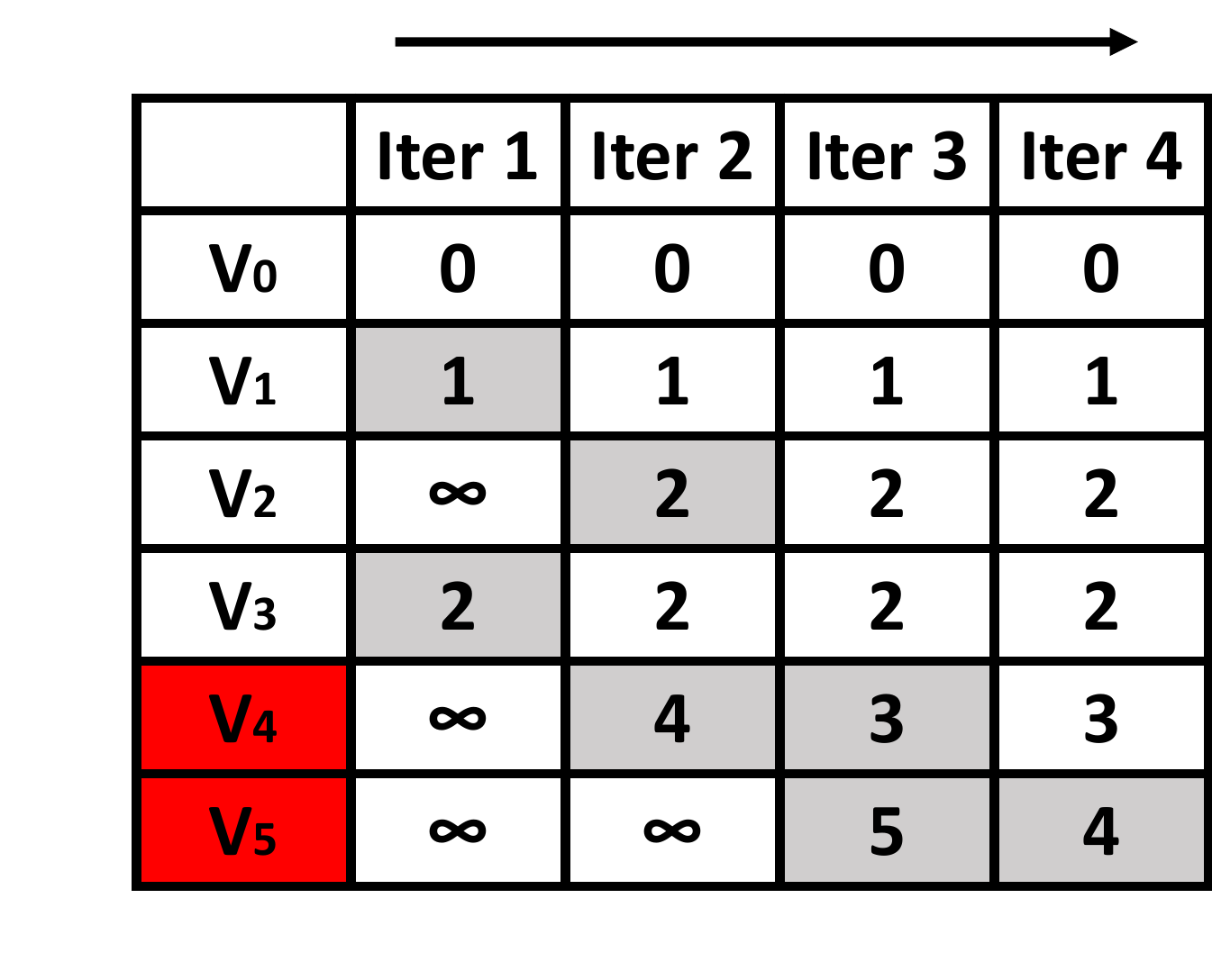}}
\end{minipage}\par\medskip
\vspace{-0.3cm}
\caption{Example of SSSP computations.}
\label{fig:motivation1}
%\vspace{-0.5cm}
\end{figure}

\begin{table}[b]
%\vspace{-0.2cm}
\renewcommand{\arraystretch}{1.3}
\caption{Updates per vertex of SSSP in PowerLyra and Gemini. ``-'' indicates failed execution. Details of OK-FS graphs are shown in Table~\ref{table:graphs}.}
\label{table:motivation1}
\centering
\scriptsize
\begin{tabular}{|c|c|c|c|c|c|c|c|} 
\hline
 & OK & LJ & WK & DI & PK & ST & FS\\
\hline
\textbf{SSSP} & & & & & & &\\
\textit{PowerLyra} & 12.4 & 8.75 & 10.3 & 6.75 & 9.25 & 7.57 & - \\
\textit{Gemini} & 9.91 & 7.66 & 7.28 & 5.6 & 9.42 & 4.51 & 8.18\\
\hline
\end{tabular}
\end{table}

Figure~\ref{fig:motivation1} demonstrates an example of SSSP execution (using $min()$ as the core computing operation) in modern graph systems.
To simplify the explanation, we denote vertex $0$ as $V_{0}$,
and an edge from vertex $0$ to $1$ as $E_{01}$.
We leverage the updates on $V_{4}$ and $V_{5}$ to demonstrate the provenance of computational redundancy.
The vertex property \textit{dist[v]} is initialized to 0 for $V_{0}$ and $\infty$ for other vertices.
During $Iter_{1}$,
the \textit{dist} of $V_{1}$ and $V_{3}$ are synchronously updated to 1 and 2,
respectively (updates are marked in gray).
In the next iteration,
the updates of $V_{1}$ and $V_{3}$ are propagated via the edges ($E_{12}$ and $E_{34}$).
Hence, the $dist$ of $V_{2}$ and $V_{4}$ are computed to 2 and 4 correspondingly.
Similarly, $V_{4}$'s property is replaced by 3 (i.e., minimum \textit{dist}) in $Iter 3$ and the \textit{dist} of $V_{5}$ updates to 5.
Due to the fact that $V_{4}$'s \textit{dist} is updated in $Iter_{3}$,
its successor---$V_{5}$'s \textit{dist} has to be recomputed in $Iter_{4}$ and updated to its minimum distance 4. 

From this example,
we can see that multiple rounds of computations are needed to calculate the shortest path for $V_{4}$ and its successor, $V_{5}$.
Such computations include multiple additions,
$min$ comparisons,
and synchronous updates,
which are time consuming in modern distributed graph systems.
Similar behaviors are observed in other graph algorithms aggregated with $min()$/$max()$ operations as well.
Such redundancies are due to the label-propagation processing manner~\cite{clip,powergraph},
where vertices are involved in computations at multiple propagation levels (e.g., $V_{4}$ resides in level 2 and 3).
Table~\ref{table:motivation1} summarizes the number of updates/computations per vertex of SSSP in PowerLyra and Gemini.
Both systems have a high number per-vertex computations,
9.1 and 7.5 on average for PowerLyra and Gemini respectively.
Note that ideally this number is 1 with no redundant computation.

\begin{figure}[t]
  \centering
  \includegraphics[height=0.1\textheight,width=0.9\linewidth, trim = 0mm 40mm 0mm 3mm, clip=true]{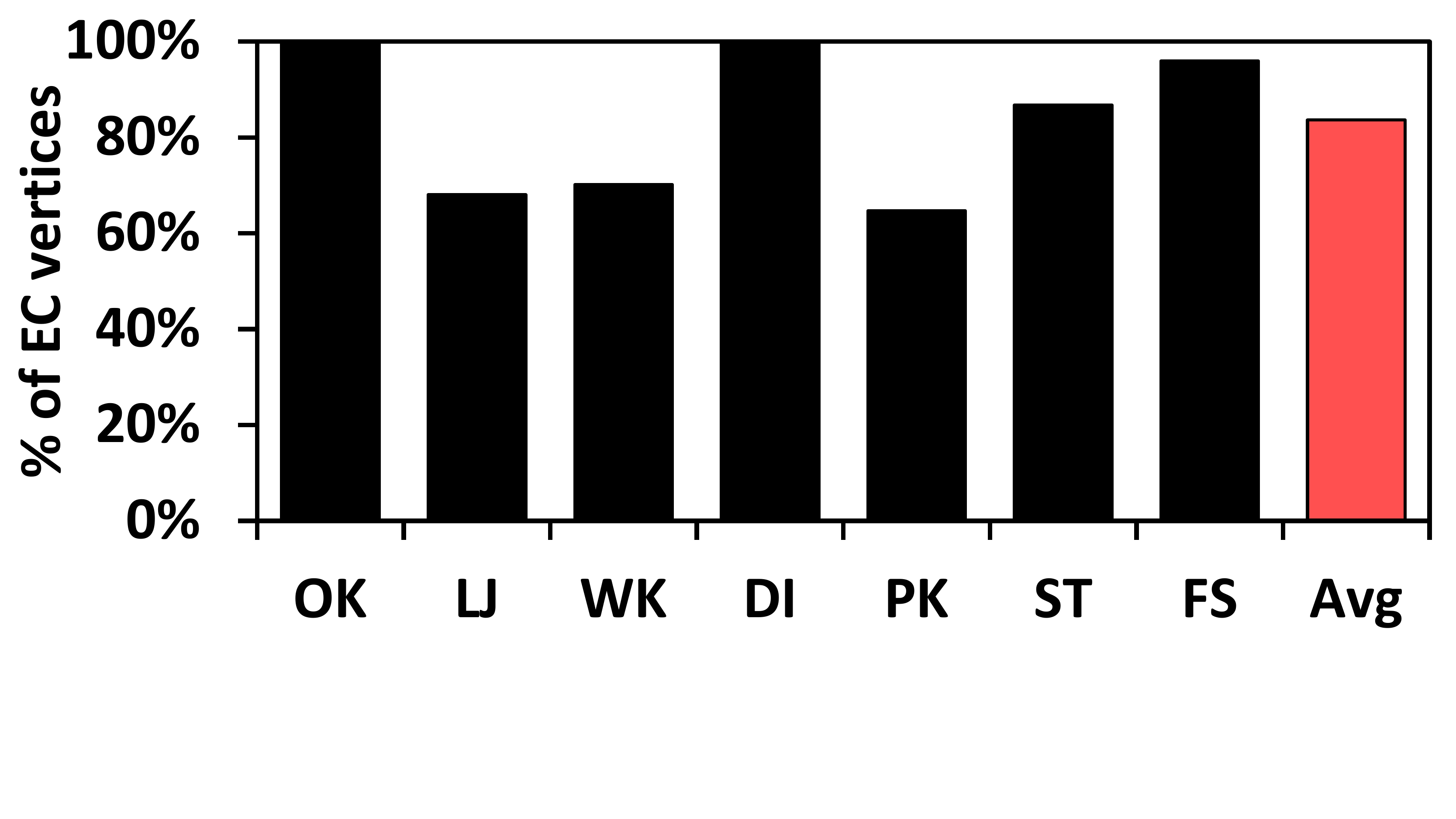} 
  \caption{Percentage of EC vertices in PR.}
  \label{fig:motivation2}
  \vspace{-0.2cm}
\end{figure}

Some other applications such as PR use arithmetic $sum()$ function for an aggregation process.
Iteratively, the values of all immediate source vertices need to be fetched for every destination vertex's computation in PR.
The convergence for this kind of algorithms is defined as the property of all vertices with no further change.
There are two reasons that a vertex's value get stabilized:
1) all the source vertices provide the same inputs as those in the past iteration;
or 2) the precision supported by the underlying hardware cannot reveal the changes.
For instance, even though the $\sum PR(v_{src})$ of two iterations are different,
dividing by the same denominator (number of links of $v_{x}$) can produce the same result.
Generally, dozens to hundreds of iterations are needed to reach a graph's stabilization.
However, we find that many vertices' properties reach a converged/stable state earlier than the entire graph's final convergence.
To quantify the percentage of these early-converged (EC) vertices,
we record every vertex's computations of seven graphs.
As shown in Figure~\ref{fig:motivation2},
a large amount of vertices have their properties stabilized,
when the program reaches 90\% of the execution time. For instance,
in OK and DI graphs,
99\% vertices are early-stabilized.
The average percentage of such vertices is 83\% across all the investigated graphs,
which indicates a room of redundancy reduction for arithmetic-based graph applications.

Even though the provenance of redundancies varies across applications,
\textit{SLFE} proposes a unified preprocessing method to generate optimization guidance for both types of redundancies. 
For $min$/$max$-based applications,
\textit{SLFE} provides a vertex's propagation sequences.
Thus, computation in all but the last iteration can be avoided (``start late''). 
For applications with arithmetic aggregation functions,
\textit{SLFE} leverages the vertex's propagation sequences to justify the status of each vertex. 
Once a vertex's property has not been updated for $x$ iterations ($x$ $>$ its maximum/latest propagation level),
the following computations on it will be avoided (``finish early'').

\section{\textit{SLFE} System Design}
\label{sec:sorre}
To eliminate redundancies for better performance, we develop \textit{SLFE}, a topology-guided distributed graph processing system, which employs the concept of ``starting late or finishing early''.
In this section, we first overview the \textit{SLFE} system, and then discuss each key technique used by \textit{SLFE} in detail.

\subsection{System Overview}
\label{sec:sysoverview}

\begin{figure}[t]
  \centering
  \includegraphics[height=0.2\textheight,width=0.975\linewidth,trim = 0mm 70mm 143mm 0mm, clip=true]{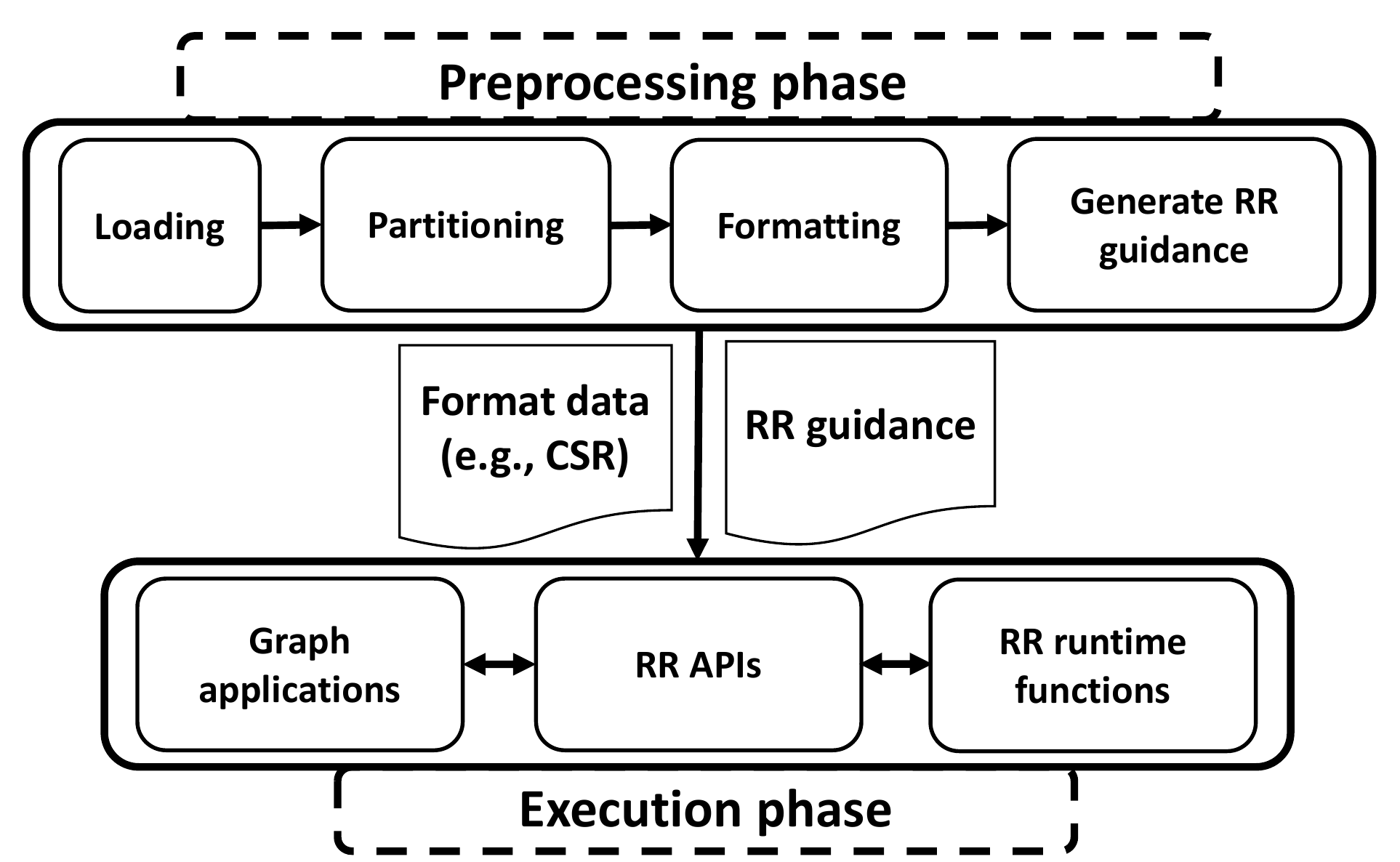} 
  \caption{System overview of \textit{SLFE}.}
  \label{fig:system_overview}
  \vspace{-0.5cm}
\end{figure}

A typical distributed graph processing system~\cite{powergraph,graphlab,pregel,geminigraph} consists of two phases: preprocessing and execution; \textit{SLFE} follows this design principle. As Figure~\ref{fig:system_overview} shows, \textit{SLFE} first loads the entire graph, partitions the graph with the fastest chunking partitioning technique available~\cite{geminigraph}, and formats the connections of the loaded graph. 
Since these components are similar to the counterparts in most existing distributed graph systems~\cite{pregel,powergraph,powerlyra,powerswitch,geminigraph}, we do not elaborate on them in this paper. Next, \textit{SLFE} performs a novel preprocessing step to generate the redundancy reduction (RR) guidance.
Such guidance indicates the  propagation sequences of an individual vertex, which \textit{SLFE} utilizes to schedule the subsequent vertex-centric operations.

In the execution phase, \textit{SLFE} follows the repeated relaxing manner to maintain a high parallelism, based on which, our vertex-centric push/pull runtime functions apply RR guidance to optimize the redundant computations on-the-fly.
Regardless of graph applications' aggregation function, \textit{SLFE}'s abstract APIs enable these applications with RR-aware computation models.
As Figure~\ref{fig:system_overview} illustrates, the three components in the execution phase interact with each other. Section~\ref{sec:rrrf} -~\ref{sec:examples} start from RR runtime functions and discuss these components in a bottom-up fashion.
Since RR could potentially introduce the workload balance due to the imbalanced redundancy reduction, \textit{SLFE} leverages a work stealing technique to address such issue. In the end, we provide a theoretical proof on the correctness of our RR technique.

\subsection{Redundancy Reduction Guidance}
\label{sec:rrg}

Algorithm~\ref{alg:preprocessing} shows the pseudo-code of the proposed preprocessing technique to generate the RR guidance (RRG). This algorithm follows the label propagation manner to record the iteration number of a vertex's last update in the process. Line 1 shows the definition of $struct\ inf$, which is used to store the RRG of each vertex with two variables --- $visited$ and $lastIter$. Line 2 initializes the $Iter$ number and declares the $dist$ array. Line 3 allocates a $struct\ inf$ array, which stores the information to guide redundancy optimization. The $fill\_source$ function in line 4 initializes all roots to 0 and other vertices to $\infty$. Starting from line 6, this procedure iterates through all the destination vertices to check whether it has an update in the current iteration. 
For all $v_{dst}$'s neighbors with incoming edges, if a neighbor's $dist$ is computed in the past round, it notifies $v_{dst}$ to update its $lastIter$ (line 9 - 10). 
This update indicates that $v_{dst}$ resides in a new propagation sequence, which occurs later than the cached one. 
Finally, line 11 - 14 calculates $v_{dst}$. 
The weight of all edges are treated as 1 (line 12), as we only need to obtain the topological knowledge of the graph. 
Only generating the topological information can expedite the preprocessing speed as well as increase the reusability of such generated information.
Moreover, we use $visited$ to only allow one computation per vertex. This is due to the fact that the first ``visit'' update $v_{dst}$ by its shortest distance, when all the edge weights are identical. This further minimizes the preprocessing overhead.
Once $v_{dst}$'s $dist$ is updated, it becomes active to propagate its value to the succeeding vertices.

\begin{algorithm}[t]
\caption{Preprocessing to Generate RR Guidance.}
\small
\begin{algorithmic}[1]
\State struct inf \{bool visited; \ uint32\_t lastIter;\};
\State int $Iter$ = 1; int $dist[NumVerts]$;
\State $inf * RRG$ = new $inf$[numV];
\State graph$\rightarrow$fill\_source($dist$); //initialize vertices
\State \textbf{for} (int $Iter$=1; active vertex exists; $Iter$++) 
\State \quad \textbf{for} $v_{dst} \in V$ 
\State \qquad \textbf{for} $v_{src} \in v_{dst}.incomingNeighbors$ 
\State \qquad\quad \textbf{if} $v_{src}.active$
\State \qquad\qquad \textbf{if} $RRG[v_{dst}]$.lastIter $<$ $Iter$ 
\State \qquad\qquad\quad $RRG[v_{dst}]$.lastIter = $Iter$;
\State \qquad\qquad \textbf{if} !$RRG[v_{dst}]$.visited 
\State \qquad\qquad\quad $dist[v_{dst}] = dist[v_{src}] + 1;$
\State \qquad\qquad\quad $RRG[v_{dst}].visited$ = true;
\State \qquad\qquad\quad $v_{dst}.active$ = true;
\end{algorithmic}
\label{alg:preprocessing}
\end{algorithm}

After Algorithm~\ref{alg:preprocessing}, each vertex maintains an ``inf''. 
$lastIter$ of each vertex indicates the last propagation level, which receives at least one update from the active source vertices. 
Any computation/update to this vertex happens before this point can be safely ignored for the redundancy reduction purpose (``start late''). In the execution phase, such information can schedule the beginning of vertex computation for an application with $min()$/$max()$ aggregation function. 
For instance, if a vertex $v_{x}$'s $lastIter$ is 3, 
all the computations happen before this iteration can be safely omitted. However, regardless of the activeness of source vertices, such strategy requires $v_{x}$ to collect the inputs from all of them to maintain the correctness.
 
Even though applications with arithmetic operations can leverage the same RRG data to remove computational redundancies, the intuition behind is different. Due to the fact that most vertices converge earlier than graph's global convergence, $lastIter$ is used to justify the status of a vertex's stability. \textit{SLFE} treats $lastIter$ as the number of iterations needed to receive any new values from source vertices. If no change occurs on a vertex's property (e.g., ranking in PR) for $x$ rounds ($x > lastIter$), it is considered as a stabilized/converged vertex. Thus, its further computations, known as redundancies, are bypassed (``finish early'' on such vertices).  

Overall, we can see that the proposed preprocessing technique is an extra step after finalizing graph partitions, which is generally applicable to any partitioning schemes and data formats. Thus, it can be easily adopted by other state-of-the-art graph systems~\cite{geminigraph,pregel,powergraph,powerlyra,powerswitch,graphchi,graphx,clip}. The overhead of our scheme is low and is thoroughly evaluated in Section~\ref{sec:eva}. In the next section, we discuss how \textit{SLFE} employs RRG in the execution phase.

\subsection{RR-aware Runtime Functions}
\label{sec:rrrf}

During the graph processing, the number of outgoing/incoming edges of active vertices in each iteration varies. 
Thus, modern graph processing systems~\cite{geminigraph,Nguyen,ligra,Beamer} leverage direction-aware propagation model---push and pull to dynamically balance the communication and computation.
This dual propagation mode significantly optimizes the graph processing procedure on-the-fly. However, such computing model increases the difficulty in applying redundancy reduction schemes at runtime. For instance, where do the redundant computations happen and how do you incorporate the generated RRG in the model? To answer these questions, we first analyze the pull/push propagation model. 

\begin{figure}[t]
  \centering
  \includegraphics[height=0.14\textheight,width=0.91\linewidth,trim = 0mm 120mm 220mm 0mm, clip=true]{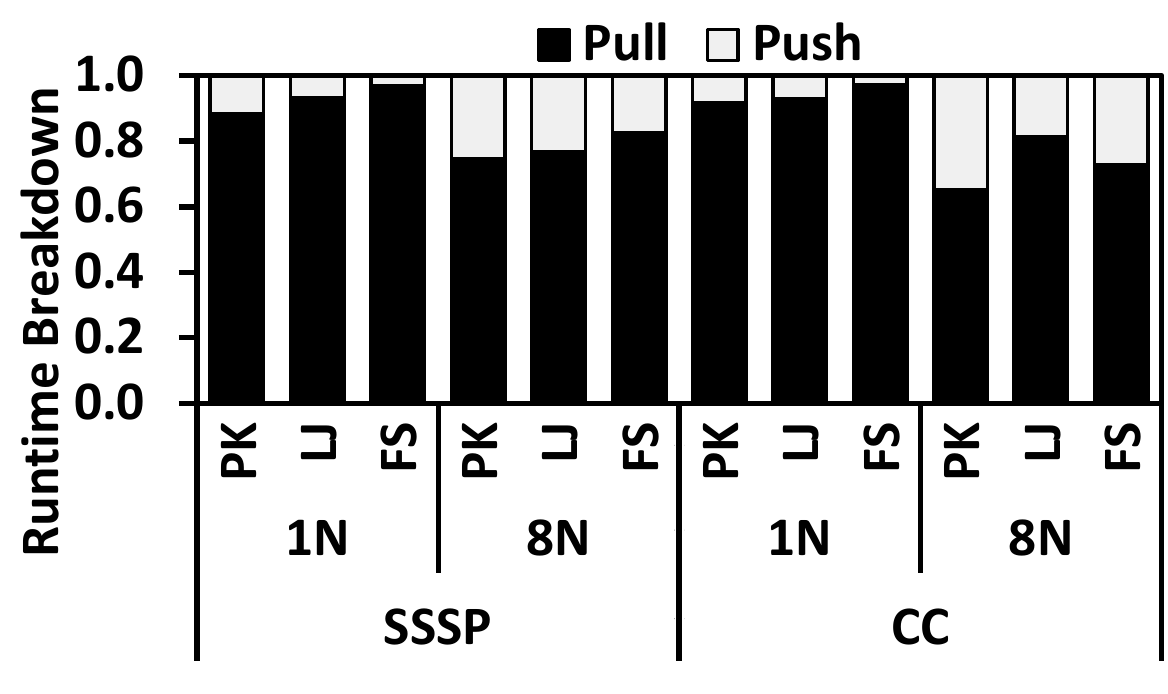} 
  \caption{SSSP and CC execution time breakdown of pull and push mode, which are measured in 1node and 8node setup with PK, LJ, and FS graphs.}
  \label{fig:pullandpush} 
  \vspace{-0.5cm}
\end{figure}

We measure the execution time of pull/push mode in SSSP and CC~\footnote{Applications with arithmetic aggregation functions always execute in the pull mode to iteratively compute all vertices, due to~\cite{activeprissue}.} with three natural graphs. The same measurements are performed in both 1 single node and 8 distributed nodes to demonstrate the increasing effect of communications (push). As Figure~\ref{fig:pullandpush} shows, SSSP and CC on a single node spend more than 92.8\% and 94\% of their execution time in the pull mode. When we run them on 8 nodes, the runtime in pull mode still consumes more than 78\% and 73\% in SSSP and CC, respectively. Such small decrement in pull is due to 1) the network advancement (e.g., the use of InfiniBand); 2) in execution, push is usually employed to kick off the execution or finish up the remaining work. We observe that most computations happen in pull, where \textit{SLFE} primarily applies the redundancy reduction scheme. We also attempt to eliminate redundancies in push. However, as the number of push operations is significantly less than the number of pull operations, the overhead of checking and removing redundancies surpasses the performance benefit. This insight is also reported in previous work~\cite{activeprissue}. Hence, rather than redundancy reductions, \textit{SLFE} leverages the push to ensure the application's correctness.  

\begin{algorithm}[t]
\caption{Pull Mode Computation.}
\small
\begin{algorithmic}[1]
\State def pullEdge\_singleRuler(pullFunc, Ruler)\{
\State \quad $pull$ = true;
\State \quad \textbf{for} $v_{dst} \in V$ \textbf{do}
\State \qquad \textbf{if} $Ruler \geq RRG[v_{dst}].lastIter$
\State \qquad\quad pullFunc($v_{dst}, v_{dst}.incomingNeighbors$);
\State \}

\State def pullEdge\_multiRuler(pullFunc, RulerS)\{
\State \quad $pull$ = true;
\State \quad \textbf{for} $v_{dst} \in V$ \textbf{do}
\State \qquad \textbf{if} $RulerS[v_{dst}] < RRG[v_{dst}].lastIter$
\State \qquad\quad pullFunc($v_{dst}, v_{dst}.incomingNeighbors$);
\State \}
\end{algorithmic}
\label{alg:pull}
\end{algorithm}

\begin{algorithm}[t]
\caption{Push Mode Computation.}
\small 
\begin{algorithmic}[1]
\State def pushEdge(pushFunc)\{
\State \quad \textbf{if} $pull$ \textbf{do} 
\State \qquad activateAllVertices();
\State \qquad $pull$ = false;
\State \quad \textbf{for} $v_{src} \in V$ \textbf{do}
\State \qquad \textbf{if} $v_{src}.hasOutgoing\ \&  \ v_{src}.active$ \textbf{do}
\State \qquad\quad pushFunc($v_{src}, v_{src}.outgoingNeighbors$);
\State \}
\end{algorithmic}
\label{alg:push}
\end{algorithm}

Algorithm~\ref{alg:pull} demonstrates the pull runtime function, which includes a $pullEdge\_singleRuler$ and a $pullEdge\_multiRuler$ function. At the beginning, variable $pull$ is set to true (we will explain the usage of this variable with push). The $pullEdge\_singleRuler$ function receives the user-defined $pullFunc$ and $Ruler$. For all $v_{dst}$s, a single $Ruler$ is used to control their executions (line 4). As aforementioned, the $lastIter$ of each vertex is used to optimize the redundancies. For instance, $min$/$max$-based algorithm uses the current iteration number as the single $Ruler$. If a vertex $v_{x}$'s $RRG.lastIter$ is 4, the beginning of its iterations will be delayed after the iteration 3. 
The $pullEdge\_multiRuler$ receives an $RulerS$ array that is transparent to users. Each vertex has its own ``Ruler'' to follow. For iterative applications with heavy arithmetic operations, $RulerS$ records each vertex's number of iterations that its property is continuously stable. Once $Rulers[v_{x}]$ passes its $lastIter$, any further computation is eliminated.   

In contrast, only one push function is shared by all applications (Algorithm~\ref{alg:push}). In line 2 - 4, it checks whether the last iteration is in Pull or not. If yes, this function activates all the vertices and sets the $pull$ back to $false$. The ``active list'' technique~\cite{pregel} is commonly deployed by modern distributed systems~\cite{galois,geminigraph,powergraph,graphlab,powergraph,powerswitch} to improve the communication efficiency. Thus, it only sends the property of active source vertices (line 6 - 7). However, due to the redundancy optimization, some ``active'' vertices may have been deactivated before reaching the push mode. Their successors may lose the opportunities to check the properties of these predecessors. Such coincidence can potentially result in correctness issues. Therefore, we need to reactivate all the vertices in the transition phase (i.e. pull$\rightarrow$push). Then, the active $v_{src}$ vertices with outgoing edges use user-defined $pushFunc$ to propagate its information.
The next section presents the APIs that are used to bridge these RR-aware push/pull computation models with graph applications.

\subsection{APIs}
\label{sec:apis}
\textit{SLFE} defines a set of application programming interfaces (APIs) in Table~\ref{table:apis}. These APIs bridge the user-defined vertex/edge operations with RR-aware runtime functions. The $edgeProc$ interface functions traverse a graph along the edges, while $vertexUpdate$ applies application-specific operations to a vertex's property. 

In the $min/max$ API, $activeVerts$ records the number of active vertices in each iteration and terminate the execution early once no active vertex exists. $Ruler$ compares with each vertex's $RRG.lastIter$ to schedule the computations. This API will be utilized for applications with $min$/$max$ aggregation functions such as SSSP. In contrast, $edgeProc$ for the $arith$ API does not need any redundancy reduction inputs from the user side. In both $edgeProc$ APIs, the number of active outgoing edges in current iteration dynamically drives the decision of using either the push or pull computation model. The $vertexUpdate$ applies user-defined $vertexFunc$ to each vertex at the end of each iteration. 

\begin{table}[t]
\renewcommand{\arraystretch}{1.3}
\caption{APIs provided by \textit{SLFE}.}
\label{table:apis}
\footnotesize
\begin{tabular}{c|c|c} 
\hline
\multicolumn{3}{l}{\textbf{min/max:} \textbf{void} edgeProc(pushFunc, pullFunc, activeVerts, Ruler);}\\
\multicolumn{3}{l}{\textbf{arith:}\ \ \textbf{void} edgeProc(pushFunc, pullFunc);} \\
\multicolumn{3}{l}{\qquad\quad \textbf{void} vertexUpdate(vertexFunc);}\\ 
\hline
\end{tabular}
\vspace{-0.5cm}
\end{table}

\subsection{Programming with \textit{SLFE}}
\label{sec:examples}
This section presents SSSP and PR applications implemented atop \textit{SLFE} as the examples to show the programmability of \textit{SLFE}. These examples show that optimizing redundant computation does not increase the programming burden, with the support of \textit{SLFE}'s APIs. 

\begin{algorithm}[t]
\caption{Single Source Shortest Path.}
\small
\begin{algorithmic}[1]
\State float\ *\ $dist$ = new float[$numV$];
\State $v_{root}.active$ = true;\ $dist[v_{root}]$ = 0.0;
\State uint32\_t $activeVerts$ = 1; uint32\_t $iter$ = 0;

\State pushFunc($v_{src}$, $v_{src}.outgoingNeighbors$)
\State \quad \textbf{for} $v_{dst} \in v_{src}.outgoingNeighbors$
\State \qquad float $newDist$ = $dist[v_{src}]$ + $v_{dst}.edgeData$;
\State \qquad \textbf{if} $newDist < dist[v_{dst}]$ 
\State \qquad\quad $dist[v_{dst}]$ = $newDist$;\ \ $v_{dst}.active$ = true;

\State pullFunc($v_{dst}$, $v_{dst}.incomingNeighbors$)
\State \quad float $miniDist$ = MAX;
\State \qquad \textbf{for} $v_{src} \in v_{dst}.incomingNeighbors$
\State \qquad\quad float $newDist$ = $dist[v_{src}]$ + $v_{src}.edgeData$;
\State \qquad\quad \textbf{if} $newDist < miniDist$
\State \qquad\qquad $miniDist$ = $newDist$;
\State \qquad\textbf{if} $dist[v_{dst}] > miniDist$
\State \qquad\quad $dist[v_{dst}]$ = $miniDist$;\ $v_{dst}.active$ = true;
\State while ($activeVerts$)
\State \qquad $slfe$.edgeProc($pushFunc$, $pullFunc$,
\State \qquad\qquad\qquad  $activeVerts$, \ $iter$++ );\ // $iter$ is Ruler
\end{algorithmic}
\label{alg:sssp}
\end{algorithm}

\paragraph{SSSP} In SSSP, a property $dist$ is attached to each vertex to store the shortest distance from the root and the destination vertex. The pseudo-code of SSSP is shown in Algorithm~\ref{alg:sssp}.
This program has to provide the user-defined $pushFunc$, $pullFunc$, activeVerts, and iteration counter (singleRuler for redundancy reduction) for \textit{SLFE} to process the active vertices along with the connected edges. In push mode (line 4 - 8), each $v_{dst}$ of $v_{src}$ will receive a $newDist$ composed by $dist[v_{src}]$ and the weight of a connected edge. To trigger such a computation, $v_{src}$ needs to be active in this iteration. If the $newDist$ is smaller than the current $dist$ of $v_{dst}$, $v_{dst}$ will be updated with this smaller value. Similarly, pull mode (line 9 - 16) iterates through a $v_{dst}$'s source vertices locally, and summarizes to get a local $miniDist$. If $miniDist$ is smaller than $dist[v_{dst}]$, then it will be sent to the machine owning $v_{dst}$ via message passing interface (MPI)~\cite{mpi}. Vertices with $dist$ updates will be activated for the next round. Once there is no active vertex anymore, the process will terminate. Clearly, compared to the SSSP implementations on other systems, our SSSP does not incur any extra effort from the programming perspective.

\begin{algorithm}[t]
\caption{PageRank.}
\small
\begin{algorithmic}[1]
\State float* $rank$ = new float[$numV$];
\State //graph traverse is similar to SSSP shown in Algorithm 4
\State //use the $edgeProc(pushFunc, pullFunc)$
\State float vOp($v_{x}$)
\State \quad $rank[v_{x}]$ = 0.15 + 0.85$*rank[v_{x}]$;
\State \quad \textbf{if} $v_{x}.hasOutgoing > 0$
\State \qquad $rank[v_{x}]$ /= $v_{x}.outEdges$;
\State \quad return $rank[v_{x}]$;
\State slfe.vertexUpdate(vOp);
\State //vertexUpdate is a system API to iterate through all $V$s
\State uint32\_t* $stableCnt$ = new uint32\_t[$numV$]; //RulerS
\State float* $stableValue$ = new float[$numV$];
\State vertexUpdate(vOp)
\State \quad \textbf{for} $v_{x} \in V$ 
\State \qquad \textbf{if} $stableCnt[v_{x}] < RRG[v_{x}].lastIter$
\State \qquad\quad float $rank$ = vOp($v_{x}$);
\State \qquad\quad \textbf{if} $rank = stableValue[v_{x}]$ \ \ $stableCnt[v_{x}]$++;
\State \qquad\quad \textbf{else} $stableCnt[v_{x}]$ = 0; $stableValue[v_{x}]$ = $rank[v_{x}]$;
\end{algorithmic}
\label{alg:pr}
\end{algorithm}

\iffalse
\begin{algorithm}
\caption{PageRank}
\small
\begin{algorithmic}[1]
\State uint32\_t* $stableCnt$ = new uint32\_t[$numV$]; //multiRuler
\State float* $stableValue$ = new float[$numV$];
\State float* $rank$ = new float[$numV$];
\State //graph traverse is similar to SSSP shown in Algorithm 4
\State //only showing vertex updates due to space limit
\State slfe.vertexUpdate(
\State \quad \textbf{if} $stableCnt[v_{x}] < slfe.RRG[v_{x}].lastIter$ \textbf{do}
\State \qquad $rank[v_{x}]$ = 0.15 + $0.85*rank[v_{x}]$;
\State \qquad \textbf{if} $rank[v_{x}] = stableValue[v_{x}]$\ \textbf{do}
\State \qquad\quad $stableCnt[v_{x}]$++;
\State \qquad \textbf{else}\ \textbf{do}
\State \qquad\quad $stableCnt[v_{x}]$ = 0;
\State \qquad\quad $stableValue[v_{x}]$ = $rank[v_{x}]$;
\State \quad \textbf{else do}
\State \qquad $rank[v_{x}]$ = $stableValue[v_{x}]$;
\State \quad \textbf{if} $v_{x}.hasOutgoing > 0$ \textbf{do}
\State \qquad $rank[v_{x}]$ /= $v_{x}.outEdges$;
\State );
\end{algorithmic}
\label{alg:pr}
\end{algorithm}
\fi

\paragraph{PageRank} Algorithm~\ref{alg:pr} shows the implementation of PR application in \textit{SLFE}. 
The $rank$ array stores the properties of all vertices. The way of programming $pushFunc$ and $pullFunc$ is very similar to SSSP shown above, so it is omitted to save space. Different from SSSP, after each iterative propagation process, PR has to apply an extra user-defined function (line 4 - 8) on vertices' aggregated properties ($rank$). PR provides such function to \textit{SLFE}'s $vertexUpdate$ API. 
The pseudo-code of $vertexUpdate$ is also shown in Algorithm~\ref{alg:pr} (line 11 - 18) to help understand how \textit{SLFE} achieves redundancy minimizations for PR like applications.
The vertex's status monitoring process happens in this function with the idea of tracking the number of continuous iterations that a vertex's $rank$ has not been changed. Every stable/converged iteration will increase the $stableCnt$ by 1 (line 17). If $v_{x}$ has a new $rank$, its $stableCnt$ will be erased and $stableValue$ will cache this new value (line 18). Once $v_{x}$'s total number of converged iterations exceeds its $RRG.lastIter$, we consider it as a early-converged (EC) vertex (line 15). Any further computation on it will be replaced by loading the cached $rank$ from $stableValue$.

These two examples show that the implementation of graph applications on \textit{SLFE} is very straightforward.
Additionally, \textit{SLFE}'s ``lazy'' redundancy reduction philosophy does not incur any heavy modification on the manner that the graph application used to be coded.
 
\subsection{Work Stealing} 
\label{sec:workstealing}
The workload balance of graph processing depends on many factors such as the initial partitioning quality, the density of active vertices on-the-fly, and so on. 
To overcome the imbalances among computing units, we follow the idea of~\cite{cilk,workstealing1,workstealing2,geminigraph} to implement a fine-grained work stealing mechanism in \textit{SLFE}. In execution, each graph is split into mini-chunks, and each mini-chunk contains 256 vertices. Such design can enhance the hardware (i.e., core and memory systems) utilization and take advantages of hardware prefetching. To minimize the overhead of stealing work, each thread only memorizes the starting point of the assigned mini-chunk, and simply uses a $for$ loop to iterate vertices in the mini-chunk.

During execution, all threads first try to finish up their originally assigned graph chunks, and then start to steal remaining tasks from the ``busy'' threads. The starting offsets and other metadata shared by threads are preserved via the atomic accesses such as $\_sync\_fetch\_and\_*$. Even though redundancy reduction may impact the workload balance across computation units, such explicit work stealing strategy can indeed solve the problem. The inter-node balance is guaranteed by the chunking-based partitioning as described in~\cite{geminigraph}. We examine the quality of inter-node workload balance in \textit{SLFE}. Results in Section~\ref{sec:eva_discuss} show that \textit{SLFE}'s redundancy-aware computation does not break the load balance achieved by the partitioning.

\subsection{Correctness Proof}
\label{sec:proof}
Most graph processing algorithms iteratively
evaluates certain nodal function $f_v$ applied at each vertex $v$.
Such function $f_v: \mathcal{V}^{(t)} \rightarrow \mathcal{V}^{(t+1)}$
takes the current value of all source vertices
$\mathcal{V}^{(t)}$ stored at iterations $t$ to produce the next state value
$\mathcal{V}^{(t+1)}$ for vertex $v$.

\begin{theorem}
  \label{th:bound}
   SSSP produced from the \textbf{delayed vertex computation} converges to the original output. 
\end{theorem}

\begin{proof}
  The nodal/aggregation function $f_v$ at each vertex in the SSSP algorithm is the $min()$ function, which is a monotonically decreasing function. The number of edges and all the edge weights are finite, therefore the value of $d_{n_k}^{(t)}$ is bounded by below as $t \rightarrow \infty$. Thus, by monotone convergence theorem~\cite{monotone}, the bypassed/delayed update procedure converges for SSSP. Moreover, since the initial graph state is the same for the original and the bypassed update procedure, these two procedures converge to the same value.
\end{proof}

If the output sequence $\{f_v(\mathcal{V}^{(0)}), \ldots, f_v(\mathcal{V}^{(t)})\}$ produced by the function $f_v$ converges as $t \rightarrow \infty$ for all $v \in \mathcal{V}$, then the output produced from the \textbf{delayed update procedure} converges to the original output $f_v(\mathcal{V}^{(t)})$ as $t \rightarrow \infty$. Similar proofs can be applied on other graph applications with monotonic behaviors. For graph applications with heavyweight arithmetic operations, \textit{SLFE} only bypass the subsequent computations on EC vertices. Thus, \textit{SLFE} always provides accurate results.

\section{Evaluation}
\label{sec:eva}
\begin{table}[t]
\renewcommand{\arraystretch}{1.3}
\caption{The real-world graph datasets~\cite{snapnets,KONECT}.}
\label{table:graphs}
\centering
\scriptsize
\begin{tabular}{|c|c|c|c|c|}
\hline 
Graph & |V| & |E| & AvgDegree & Type \\
\hline
pokec (PK) & 1.6M & 30.6M & 18.8 & Social\\
\hline
orkut (OK) & 3.1M & 117.2M & 38.1 &Social\\
\hline
livejournal (LJ) & 4.8M & 69M & 14.23 & Social\\
\hline
wiki (WK) & 12.1M & 378.1M & 31.1 & Hyperlink\\
\hline
delicious (DI) & 33.8M & 301.2M & 8.9 & Folksonomy\\
\hline
s-twitter (ST) & 11.3M & 85.3M & 7.5 & Social\\
\hline
friendster (FS) & 65.6M & 1.8B & 27.5 & Social\\
\hline
Synthetic (RMAT) & 300M & 10B & 33.3 & RMAT\\
\hline 
\end{tabular}
\vspace{-0.3cm}
\end{table}

\begin{table}[t]
\renewcommand{\arraystretch}{1.3}
\caption{8-node runtime (in seconds) and improvement of \textit{SLFE} over the state-of-the-art distributed systems. The per-iteration runtime is reported for PR and TR.}
\label{table:perfPGPL}
\centering
\scriptsize
\begin{tabular}{|c|c|c|c|c|c|c|c|} 
\hline
& PK & OK & LJ & WK & DI & ST & FS\\
\hline
\textbf{SSSP} & & & & & & & \\
PowerG &12.9&34.2&27.5&69.9&78.4&24.5&511\\
PowerL &10.3&23.0&18.8&34.5&18.9&17.3&243\\
\textit{SLFE} &0.58&2.5&3.98&2.8&3.1&2.3&6.25\\
\hline
\textit{Speedup}($\times$) &19.8&11.2&5.7&17.4&12.4&8.9&56.4\\
\hline
\textbf{CC} & & & & & & & \\
PowerG &7.1&19.4&15.1&26.7&47.6&14.3&236\\
PowerL &5.7&10.4&10.8&15.6&14.2&3.0&112\\
\textit{SLFE} &0.39&0.19&0.45&0.52&0.8&0.46&3.06\\
\hline
\textit{Speedup}($\times$)&16.2&74.8&28.4&39.2&32.5&14.2&53.2\\
\hline
\textbf{WP} & & & & & & & \\
PowerG &7.0&15.5&19.8&47.8&29.4&7.0&299\\
PowerL &6.1&10.2&16.0&33.1&11.1&5.3&164\\
\textit{SLFE} &0.33&0.87&0.65&0.84&2.4&0.69&3.78\\
\hline
\textit{Speedup}($\times$) &19.8&14.5&27.4&47.3&7.5&8.8&58.5\\
\hline
\textbf{PR} & & & & & & & \\
PowerG &0.71&2.20&2.10&4.05&8.67&2.01&19.2\\
PowerL &0.44&0.82&0.77&1.61&1.14&0.42&9.44\\
\textit{SLFE} &0.02&0.024&0.025&0.06&0.078&0.032&0.25\\
\hline
\textit{Speedup}($\times$) &28.0&56.0&59.9&42.6&40.3&28.7&53.8\\
\hline
\textbf{TR} & & & & & & & \\
PowerG &0.73&1.86&1.66&2.92&4.50&1.92&13.5\\
PowerL &0.28&0.69&0.74&1.65&1.11&0.37&6.07\\
\textit{SLFE} &0.05&0.02&0.04&0.05&0.1&0.03&0.34\\
\hline
\textit{Speedup}($\times$) &9.04&56.6&27.7&43.9&22.4&28.1&26.6\\
\hline
\textit{GEOMEAN} & \multicolumn{7}{|c|}{25.39$\times$}\\
\hline 
\end{tabular}
\end{table}

This section presents our evaluation results on \textit{SLFE} and compares its runtime with state-of-the-art distributed graph processing systems Gemini, PowerGraph, and PowerLyra. In addition, we also compare the single-node performance with GraphChi and Ligra, as GraphChi has been used as a baseline for many shared-memory processing systems~\cite{clip,rajivhpdc,rajiv1,xstream,gridgraph} and Ligra has superior performance compared with single-node executions of most distributed systems.
Moreover, we report \textit{SLFE}'s scalability, overhead of RRG generation, computation reduction, and inter/intra-node imbalance.   

\subsection{Experiment Setup}
We perform the experiments on the servers equipped with 2nd generation of Xeon Phi 7250 processor (Knights Landing) and InfiniBand switch (up to 100Gb/s). Each machine has 68 physical cores. Each core has a 32KB L1 I/D cache. A pair of cores share a 1MB L2 cache. HMC is deployed 
as the memory subsystem (96GB DDR4 DRAM and 16GB MCDRAM configured as the LLC). Each machine runs CentOS 7.

We use five popular graph applications from the two categories ($min$/$max$: SSSP, ConnectedComponents(CC), WidestPath (WP); $arithmetic$: PR and TunkRank(TR)). In addition, we deploy seven real-world graphs and one synthetic RMAT graph~\cite{parmat} with the number of vertices ranging from 1.6 to 300 millions and the number of edges ranging from 30 million to 10 billion (details summarized in Table~\ref{table:graphs}). Considering our cluster size (up to 8 nodes), the trillion-edge graph (around 8,000GB)~\cite{mosaic} is  inappropriate for the evaluation of \textit{SLFE}, which is not an out-of-core system.

\subsection{Overall Performance}

As \textit{SLFE} aims to reduce computational redundancies for distributed graph processing systems, comparing the runtime to other state-of-the-art distributed systems can help quantify \textit{SLFE}'s computational efficiency and high performance improvement. Table~\ref{table:perfPGPL} reports the 8-node performance of PowerGraph, PowerLyra and \textit{SLFE}, running five popular applications on seven real graphs. The results show that \textit{SLFE} outperforms these two systems in all cases significantly (25.39$\times$ on average), with up to 74.8$\times$ for CC on the OK graph. For the FS graph with billion edges, \textit{SLFE} achieves the highest average speedup (47.9$\times$) among all input datasets. Hence, in contrast to these in-memory distributed systems, \textit{SLFE} can handle large graphs more efficiently. 

\begin{figure}[t]
  \centering
  \includegraphics[height=0.125\textheight,width=1\linewidth,trim = 0mm 56mm 0mm 0mm, clip=true]{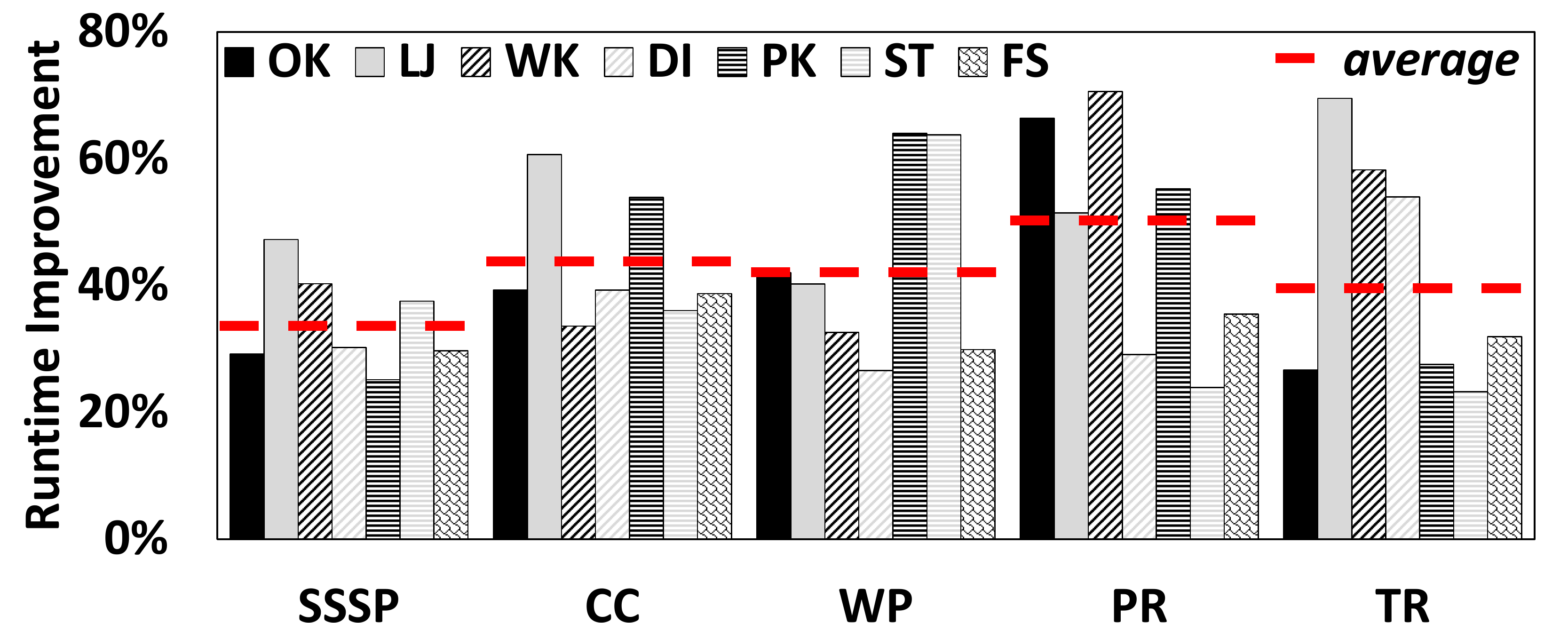} 
  \caption{\textit{SLFE}'s runtime improvement over Gemini on a 8-node cluster.}
  \label{fig:perfgemini}
  \vspace{-0.3cm}
\end{figure}

\begin{figure}[t]
\begin{minipage}{0.25\textwidth}
 \subfloat[CC-FS]{\label{fig:scaleup_cc_fs}\includegraphics[width=1\linewidth,height=0.1\textheight,trim = 3mm 130mm 230mm 0mm, clip=true]{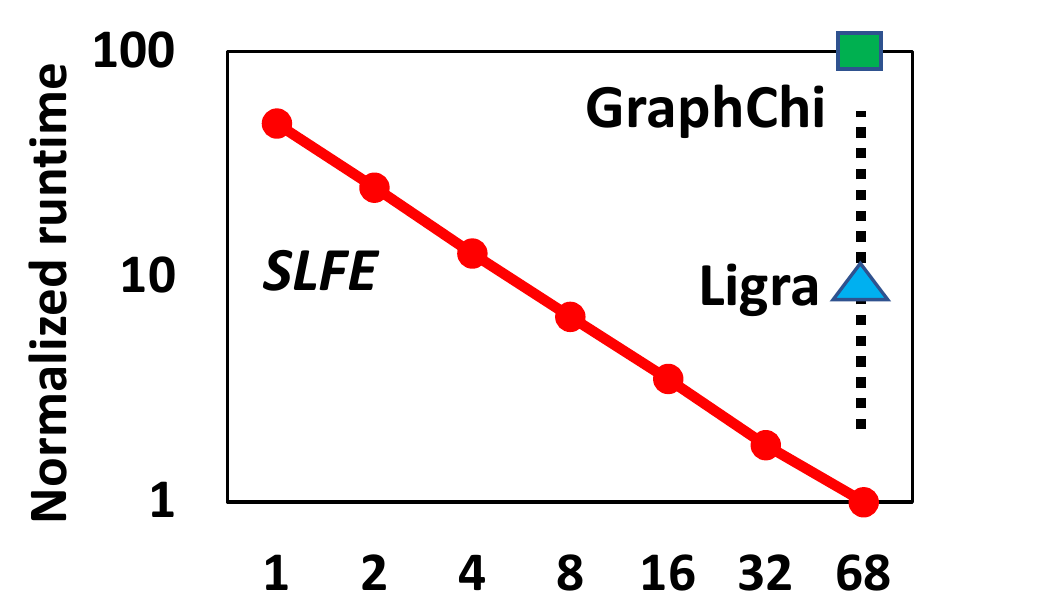}}
\end{minipage}%
\begin{minipage}{0.25\textwidth}
 \subfloat[CC-LJ]{\label{fig:scaleup_cc_lj}\includegraphics[width=1\linewidth,height=0.1\textheight,trim = 0mm 130mm 230mm 0mm, clip=true]{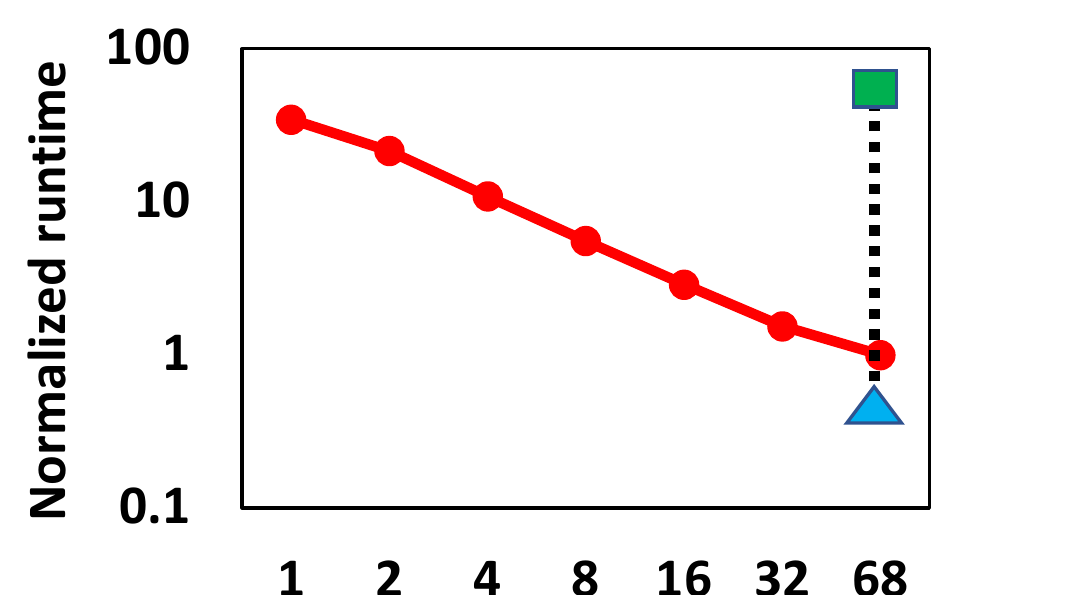}}
\end{minipage}%

\begin{minipage}{0.25\textwidth}
 \subfloat[PageRank-FS]{\label{fig:scaleup_pr_fs}\includegraphics[width=1\linewidth,height=0.1\textheight,trim = 0mm 130mm 230mm 0mm, clip=true]{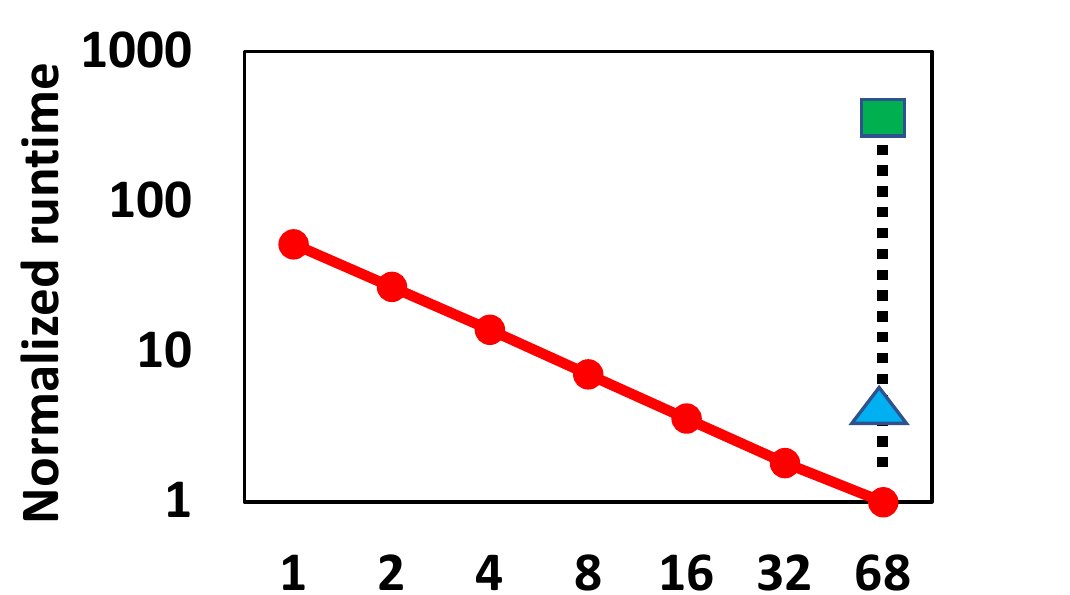}}
\end{minipage}%
\begin{minipage}{0.25\textwidth}
 \subfloat[PageRank-LJ]{\label{fig:scaleup_pr_lj}\includegraphics[width=1\linewidth,height=0.1\textheight,trim = 0mm 130mm 230mm 0mm, clip=true]{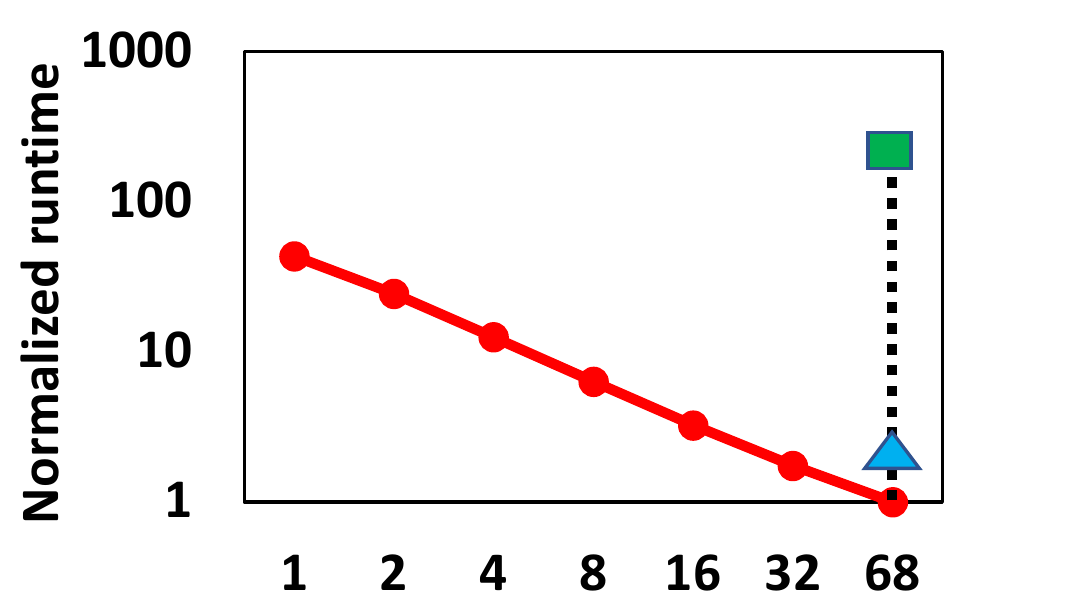}}
\end{minipage}
\caption{Intra-node scalability of \textit{SLFE}(1 - 68 cores).}
\label{fig:scaleup}
\vspace{-0.5cm}
\end{figure}

While PowerGraph and PowerLyra are the general distributed graph platforms that provide many options for designers to test their ideas, Gemini is a dedicated computation-centric system that utilizes most of the state-of-the-art optimization techniques. In~\cite{geminigraph}, Gemini has a competitive single-node performance compared to Ligra~\cite{ligra} and Galois~\cite{galois}, and outperforms PowerGraph, GraphX, and PowerLyra by 19$\times$ on average.
We compare \textit{SLFE}'s performance with Gemini in Figure~\ref{fig:perfgemini}. On average, across all seven graphs, \textit{SLFE} outperforms Gemini by 34.2\%, 43.1\%, 42.7\%, 47.5\% and 41.6\% on SSSP, CC, WP, PR, and TR, respectively. These performance gains show the effectiveness of \textit{SLFE}'s unique redundancy optimization.

Overall, \textit{SLFE}'s significant speedup comes from the reduction of redundant computation and communication. For distributed graph processing, the update on a vertex triggers either a local atomic operation or a remote synchronization via the Ethernet. Due to the computation-centric design of Gemini, the overhead of communication is not well managed compared to PowerLyra and PowerGraph. Hence, as the cluster size increases, the communication overhead surpasses the benefits obtained from adding more computation resources (Figure~\ref{fig:scaleout_pr_wk}). 
In contrast, \textit{SLFE} reduces the number of computations, resulting in fewer updates, and thus less communication across distributed machines. 
Such benefits can be observed on relatively smaller graphs such as OK, LJ, and WK, where communication effect is amplified (up to 71\% improvement). For the large FS graph, \textit{SLFE} outperforms Gemini in all applications by 33.2\% on average. Such improvement is mainly from the optimization of redundant computation, which dominates the execution time.

\subsection{Scalability}

\begin{figure}[t]
\begin{minipage}{0.25\textwidth}
\centering
 \subfloat[PageRank-FS]{\label{fig:scaleout_pr_fs}\includegraphics[width=0.9\linewidth,height=0.09\textheight,trim = 0mm 110mm 215mm 0mm, clip=true]{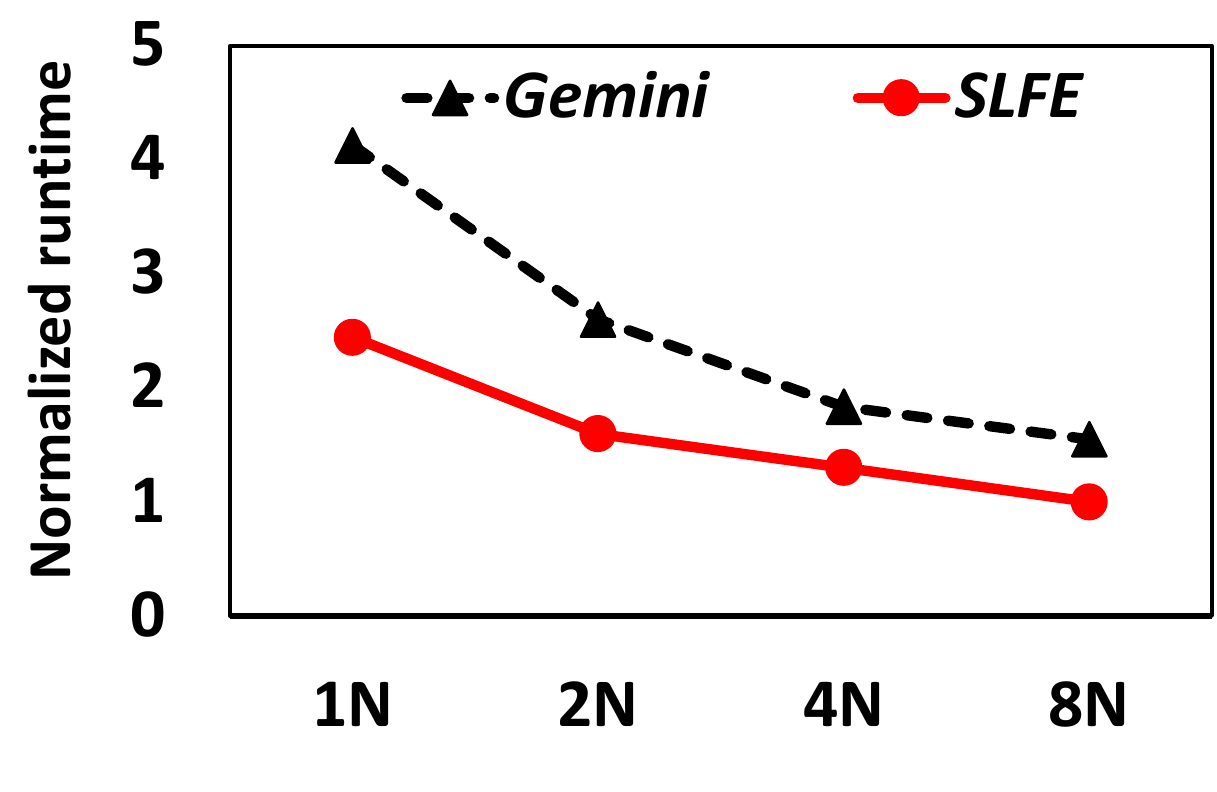}}
\end{minipage}%
\begin{minipage}{0.25\textwidth}
\centering
 \subfloat[PageRank-WK]{\label{fig:scaleout_pr_wk}\includegraphics[width=0.9\linewidth,height=0.09\textheight,trim = 0mm 110mm 215mm 0mm, clip=true]{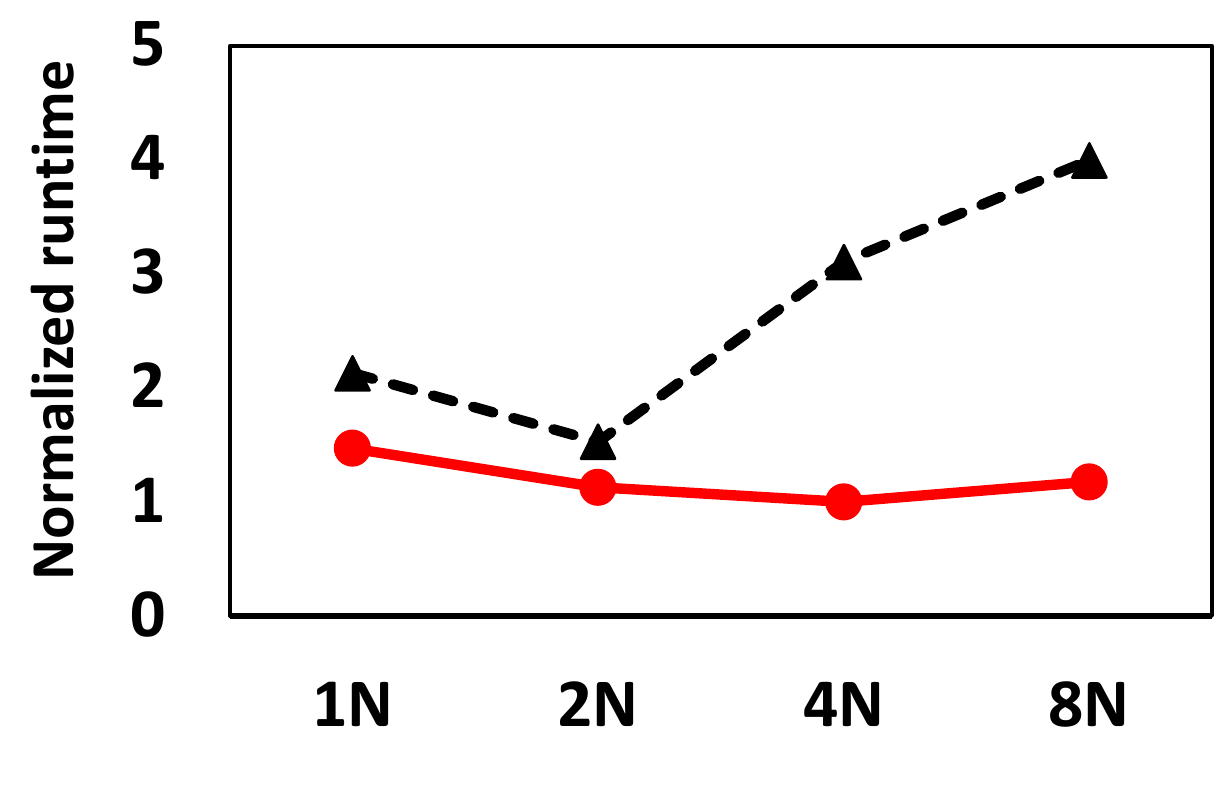}}
\end{minipage}\par\medskip

\begin{minipage}{0.25\textwidth}
\centering
 \subfloat[CC-FS]{\label{fig:scaleout_cc_fs}\includegraphics[width=0.9\linewidth,height=0.09\textheight,trim = 0mm 110mm 215mm 0mm, clip=true]{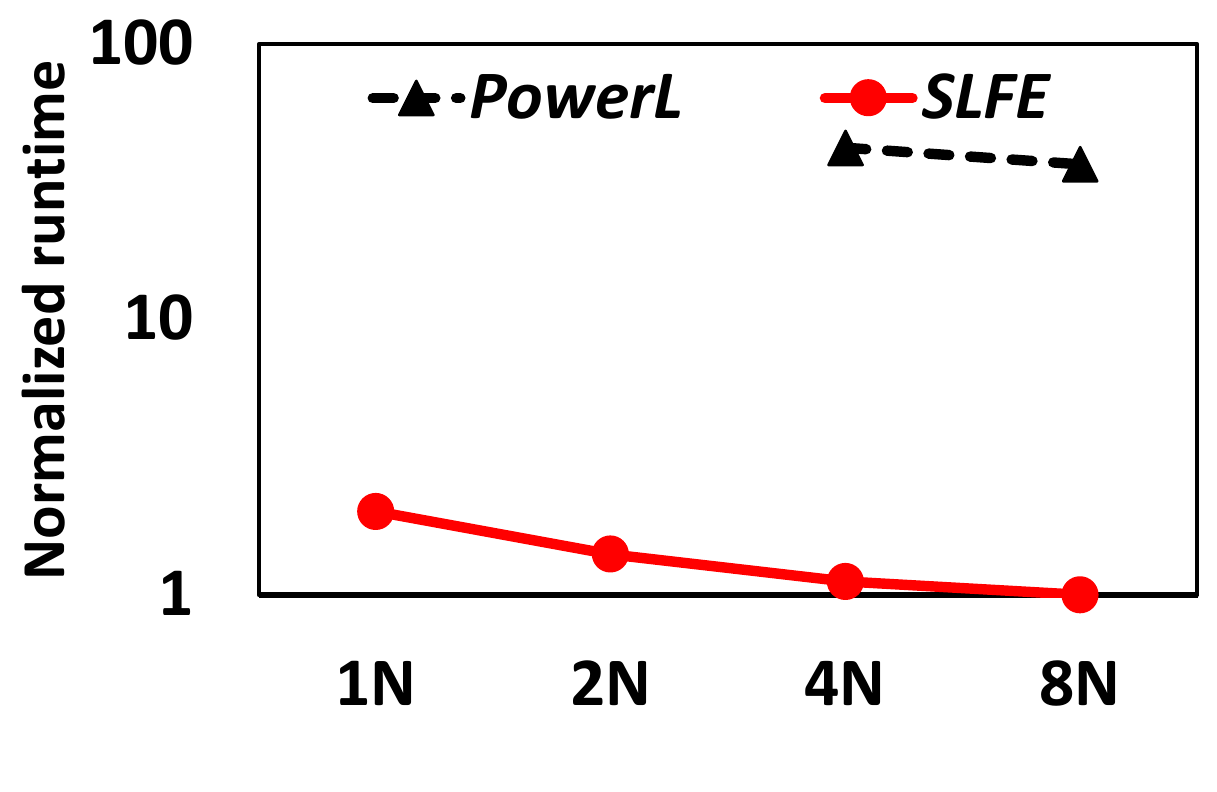}}
\end{minipage}%
\begin{minipage}{0.25\textwidth}
\centering 
\subfloat[CC-WK]{\label{fig:scaleout_cc_wk}\includegraphics[width=0.9\linewidth,height=0.09\textheight,trim = 0mm 110mm 215mm 0mm, clip=true]{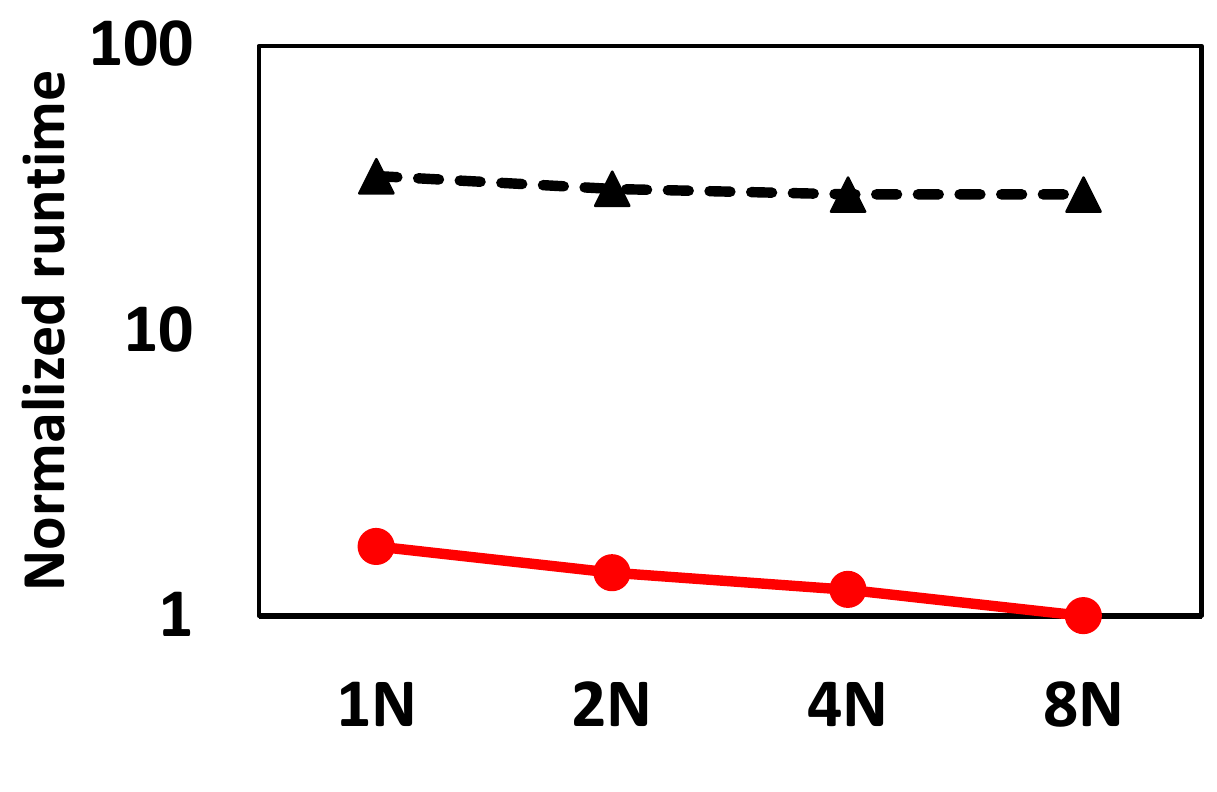}}
\end{minipage}\par\medskip

\begin{minipage}{0.5\textwidth}
\centering
 \subfloat[\textit{SLFE} with RMAT]
 {\label{fig:scaleout_rmat}\includegraphics[width=0.85\linewidth,height=0.12\textheight,trim = 0mm 132mm 240mm 0mm, clip=true]{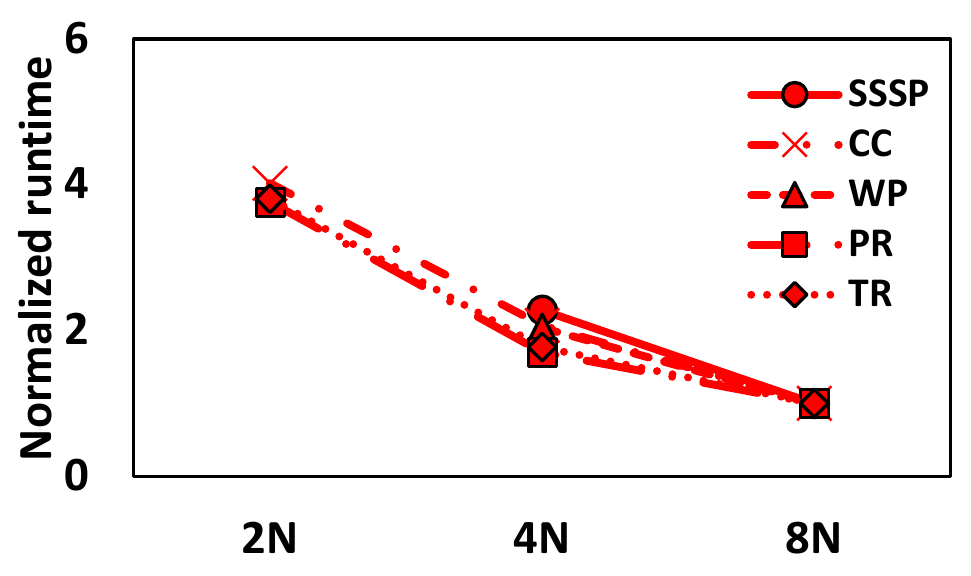}}
\end{minipage}\par\medskip
\vspace{-0.2cm}
\caption{Inter-node scalability of PowerLyra, Gemini, and \textit{SLFE} (1 - 8 nodes).}
\label{fig:scaleout}
\vspace{-0.6cm}
\end{figure} 

Next, we examine the scalability of \textit{SLFE}, starting from the intra-node experiments using 1 to 68 cores to run CC and PR on the FS and LJ graphs. Overall, Figure~\ref{fig:scaleup} shows that \textit{SLFE} achieves nearly linear scale-up in all cases. For instance, compared to 1-core and 32-core cases, running on 68 cores achieves an average speedup of 45$\times$ and 1.73$\times$, respectively.  
Although the pressure on shared hardware resources becomes more intensive as core count goes up, \textit{SLFE} still maintains a decent speedup curve.

Moreover, we also compare \textit{SLFE}'s performance with two state-of-the-art single machine systems---GraphChi and Ligra. 
\textit{SLFE} achieves up to 508$\times$ and 9.3$\times$ speedup over GraphChi (Figure~\ref{fig:scaleup_pr_fs}) and Ligra (Figure~\ref{fig:scaleup_cc_fs}), respectively. GraphChi uses cost-efficiency to trade-off performance, where its bottleneck is the intensive I/O accesses. In contrast, Ligra takes the advantages of processing entire graph loaded in the memory. However, it produces excessive amount of computations and memory accesses. \textit{SLFE} reduces the computational redundancies, and then results in less CPU usage and memory accesses in the shared-memory platform. 

PowerLyra and Gemini are used as the baselines to demonstrate \textit{SLFE}'s inter-node scalability. Figure~\ref{fig:scaleout_pr_fs} and~\ref{fig:scaleout_pr_wk} show that \textit{SLFE} constantly outperforms Gemini in all cluster configurations. In addition, when Gemini executes PR with WK graph, there is an inflection point and its performance starts to drop when running on 2 nodes. In contrast, as \textit{SLFE} eliminates redundant computations on EC vertices, the negative impact of communication in the large cluster have been reduced. Compared to PowerLyra on CC, \textit{SLFE} has a better scaling trend (Figure~\ref{fig:scaleout_cc_fs} and~\ref{fig:scaleout_cc_wk}). 

However, due to the limited size of the real graphs we have, it is difficult to demonstrate \textit{SLFE}'s scale-out capability. Thus, we generate a synthetic RMAT graph~\cite{parmat}, which has 300 million vertices and 10 billion edges. This graph exceeds the memory capacity of a single node, so we run it starting from 2 nodes. We utilize this large graph to further verify \textit{SLFE}'s scalability. 
Figure~\ref{fig:scaleout_rmat} shows that running on 8 nodes achieves 3.85$\times$ and 1.96$\times$ speedup over 2 and 4 nodes, respectively.
Such scale-out trends verify \textit{SLFE}'s design philosophy of optimizing redundancies while maintaining high parallelism that has been achieved by existing graph processing techniques. 

\begin{figure}[t]
  \centering
  \includegraphics[height=0.125\textheight,width=0.95\linewidth,trim = 0mm 125mm 210mm 0mm, clip=true]{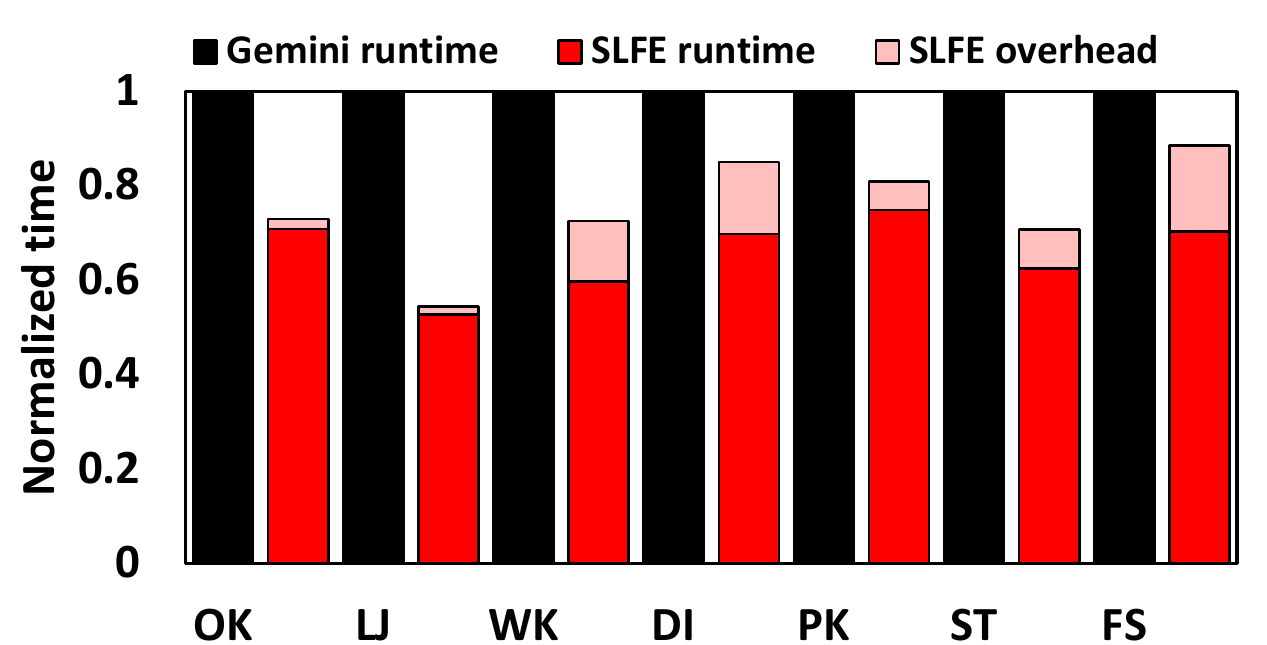} 
  \vspace{-0.1cm}
  \caption{Preprocessing overhead analysis on SSSP.}
  \label{fig:overhead}
  \vspace{-0.5cm}
\end{figure}

\subsection{Overhead Analysis}
Compared to modern distributed graph processing systems, \textit{SLFE}'s RRG generation is the only extra step. Since Gemini is one of the fastest distributed systems, we use Gemini with ``active list'' technique as the baseline here to report \textit{SLFE}'s overheads. Among all the five applications in Figure~\ref{fig:perfgemini}, \textit{SLFE} achieves the lowest performance improvement in SSSP. 
Hence, we compare \textit{SLFE}'s execution time on SSSP and the preprocessing cost to Gemini's sole runtime~\footnote{Partitioning time is not included, as \textit{SLFE} uses the same Chunking-based partitioning as Gemini.} in Figure~\ref{fig:overhead}. Such cost is extremely small on the relatively small graphs such as OK, LJ, and PK. As the graph size increases, the overhead proportionally gets slightly larger (e.g., WK and FS). However, even including our preprocessing cost, \textit{SLFE} can achieve an average of 25.1\% ``end-to-end'' performance improvement over Gemini. 
Additionally, the generated guidance can be used repeatedly by many applications to save the preprocessing cost~\footnote{In practice, Facebook~\cite{facebook} uses Giraph~\cite{giraph} to handle the applications such as the variants of PageRank and SSSP on the same graph. The average number of jobs applied to the same graph is 8.7.}. 
Therefore, \textit{SLFE}'s low overhead is acceptable and can be easily amortized.

\begin{figure}[t]
\begin{minipage}{0.25\textwidth}
 \subfloat[SSSP-FS]{\label{fig:iterplot_sssp_fs}\includegraphics[width=1\linewidth,height=0.1\textheight,trim = 0mm 109mm 210mm 0mm, clip=true]{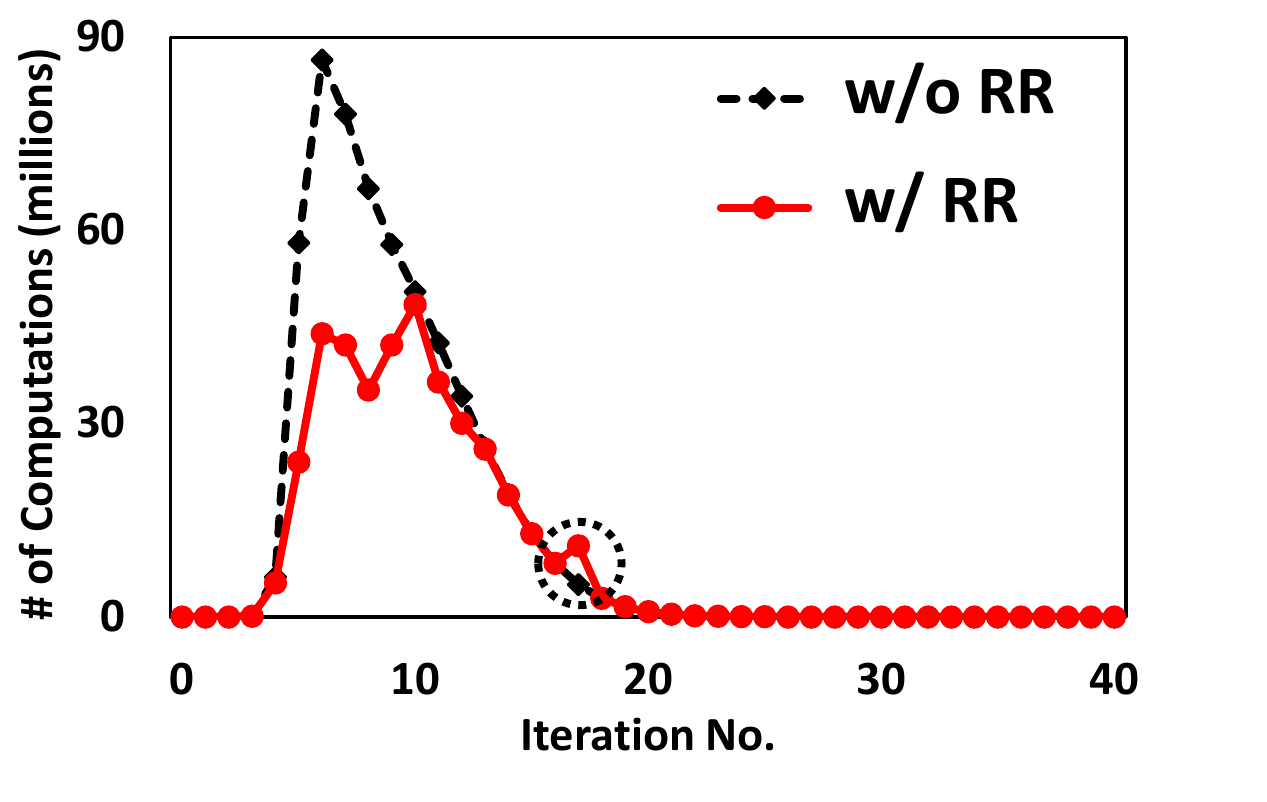}}
\end{minipage}%
\begin{minipage}{0.25\textwidth}
 \subfloat[SSSP-LJ]{\label{fig:iterplot_sssp_lj}\includegraphics[width=1\linewidth,height=0.1\textheight,trim = 0mm 109mm 210mm 0mm, clip=true]{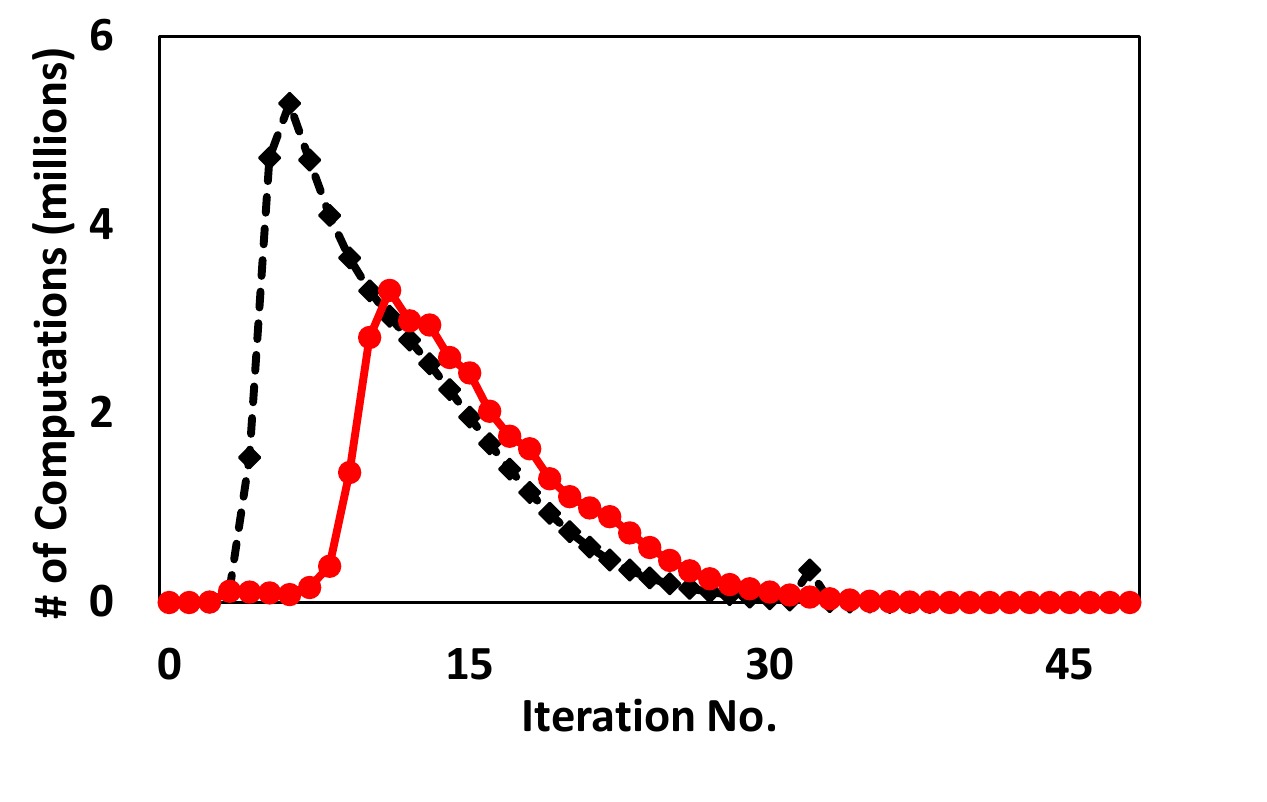}}
\end{minipage}%

\begin{minipage}{0.25\textwidth}
 \subfloat[CC-FS]{\label{fig:iterplot_cc_fs}\includegraphics[width=1\linewidth,height=0.1\textheight,trim = 0mm 109mm 210mm 0mm, clip=true]{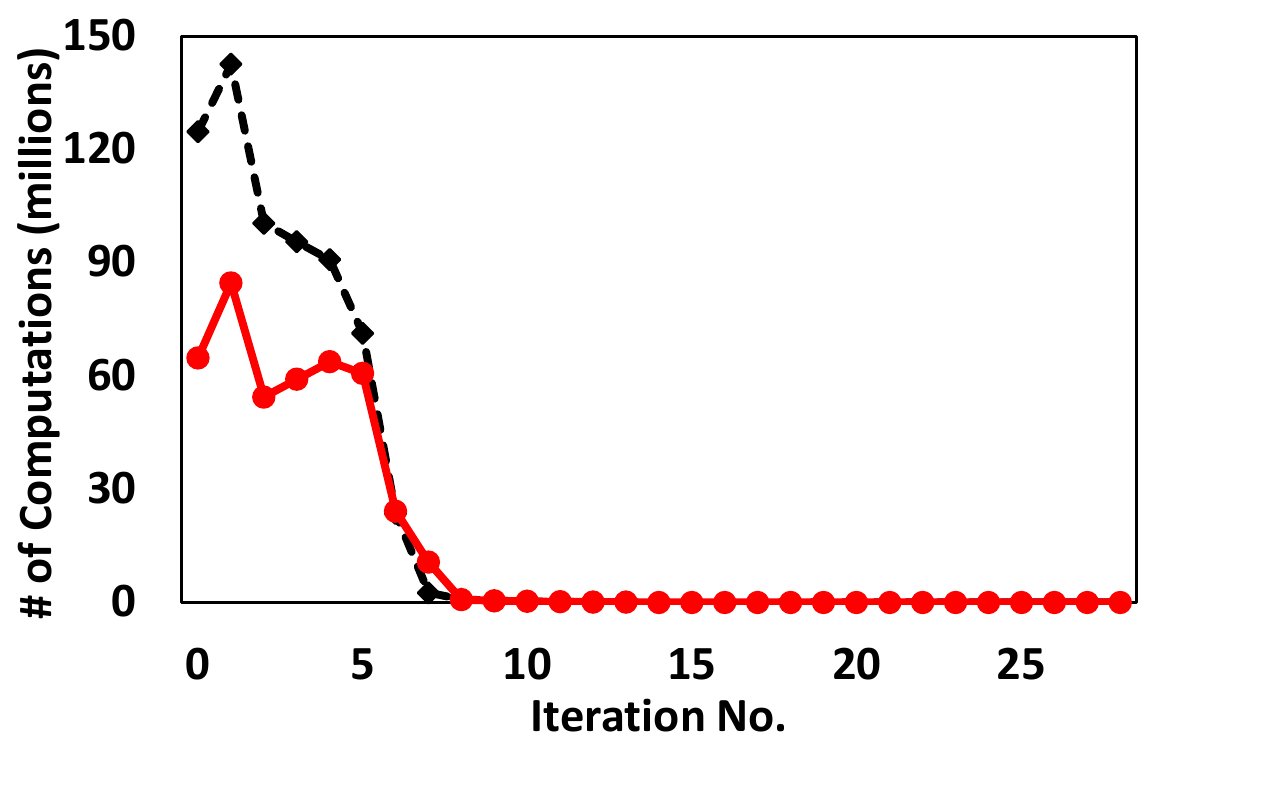}}
\end{minipage}%
\begin{minipage}{0.25\textwidth}
 \subfloat[CC-LJ]{\label{fig:iterplot_cc_lj}\includegraphics[width=1\linewidth,height=0.1\textheight,trim = 0mm 109mm 210mm 0mm, clip=true]{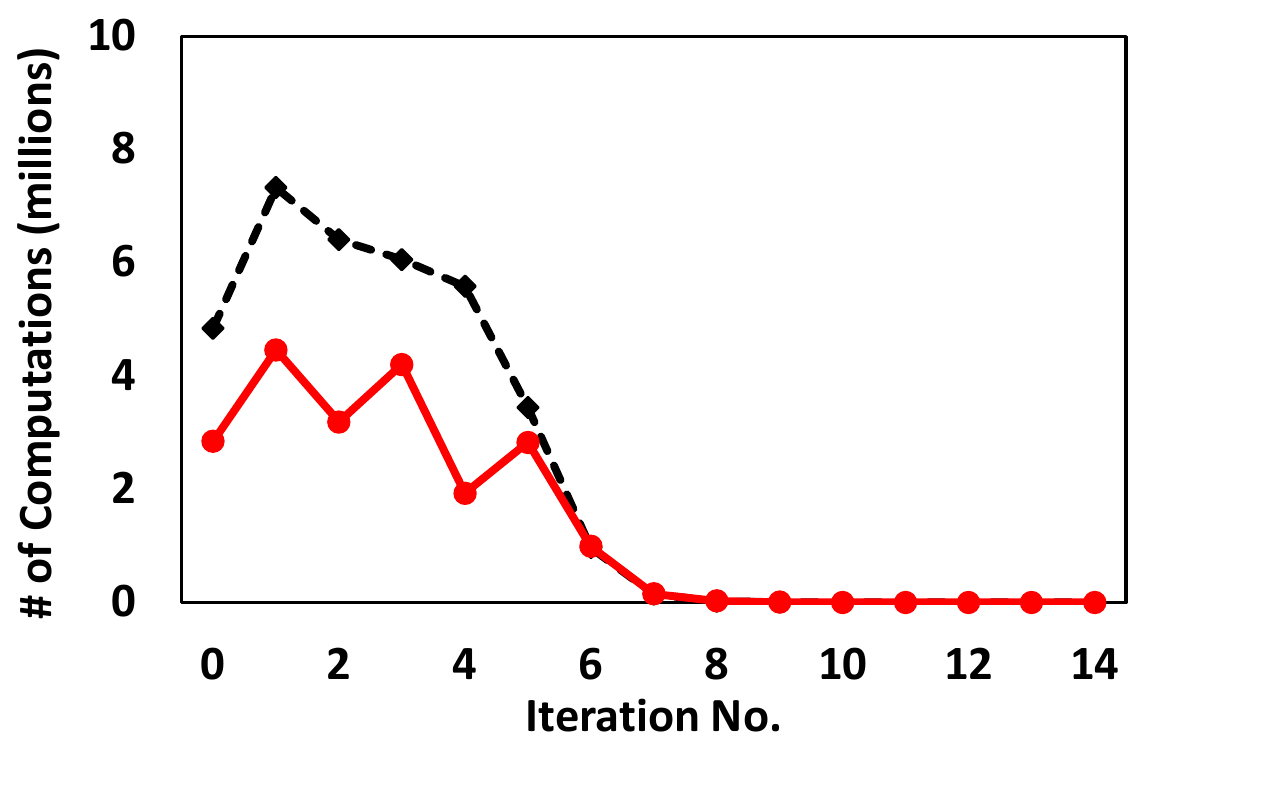}}
\end{minipage}%

\begin{minipage}{0.25\textwidth}
 \subfloat[PageRank-FS]{\label{fig:iterplot_pr_fs}\includegraphics[width=1\linewidth,height=0.1\textheight,trim = 0mm 109mm 210mm 0mm, clip=true]{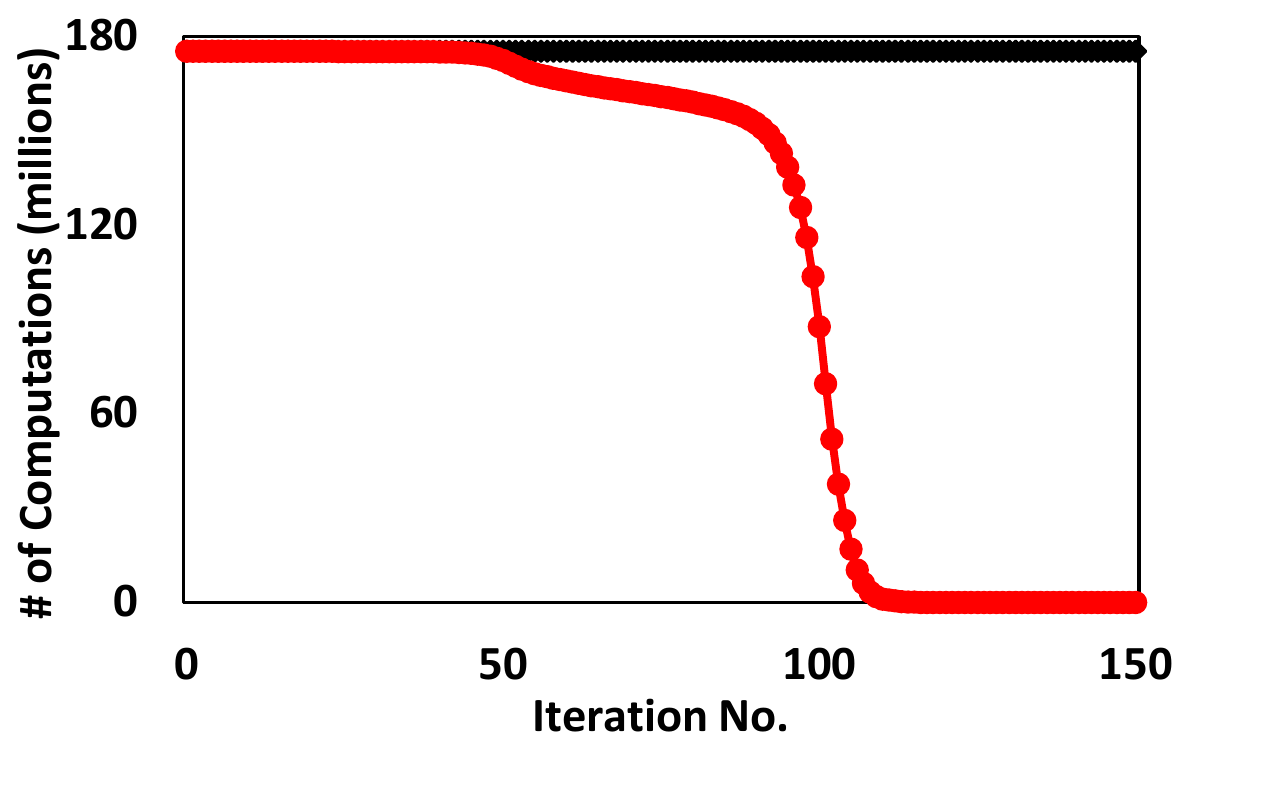}}
\end{minipage}%
\begin{minipage}{0.25\textwidth}
 \subfloat[PageRank-LJ]{\label{fig:iterplot_pr_lj}\includegraphics[width=1\linewidth,height=0.1\textheight,trim = 0mm 109mm 210mm 0mm, clip=true]{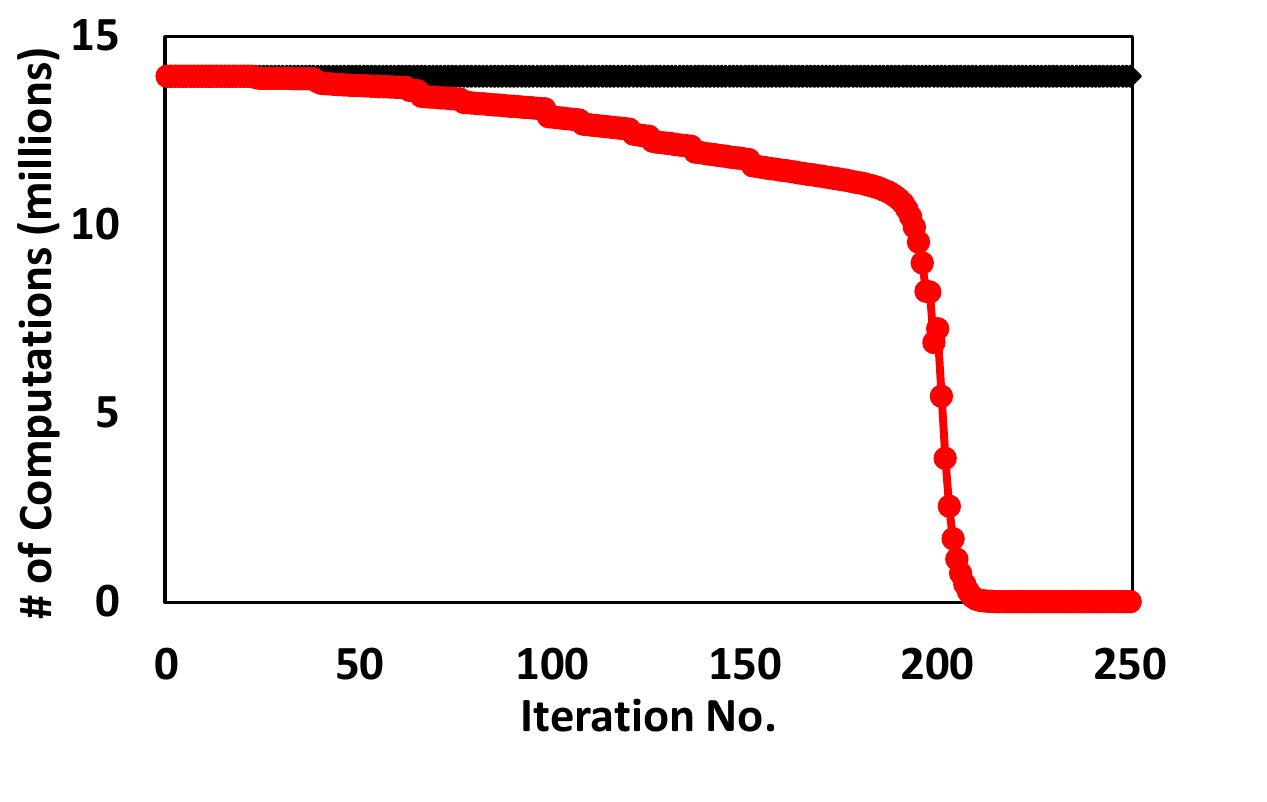}}
\end{minipage}%
\vspace{-0.1cm}
\caption{\textit{SLFE}'s no. of computations per iteration.}
\label{fig:iterplot}
\end{figure}

\begin{figure}[t]
\begin{minipage}{0.25\textwidth}
 \subfloat[Intra-node imbalance]{\label{fig:intranode}\includegraphics[width=1\linewidth,height=0.1\textheight,trim = 0mm 135mm 240mm 0mm, clip=true]{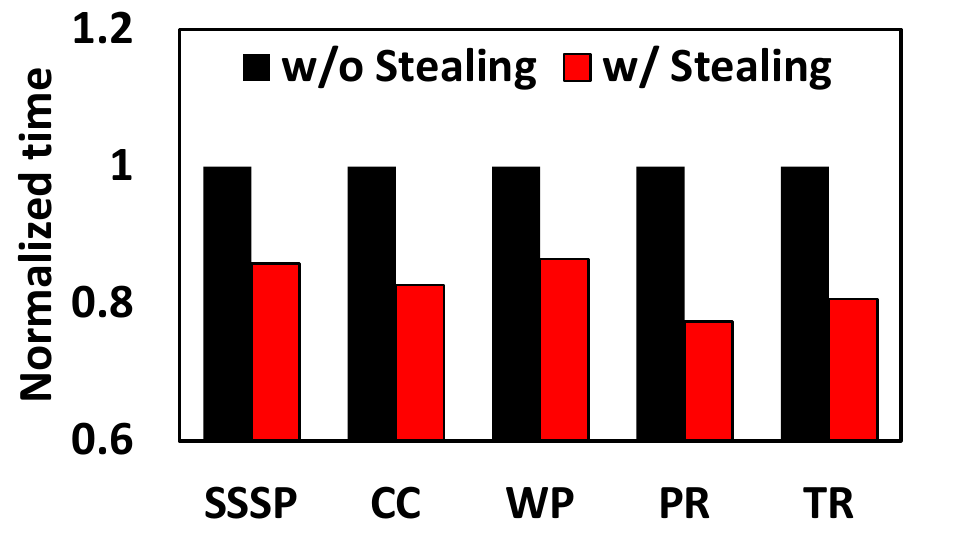}}
\end{minipage}%
\begin{minipage}{0.25\textwidth}
 \subfloat[Inter-node imbalance]{\label{fig:internode}\includegraphics[width=1\linewidth,height=0.1\textheight,trim = 0mm 135mm 240mm 0mm, clip=true]{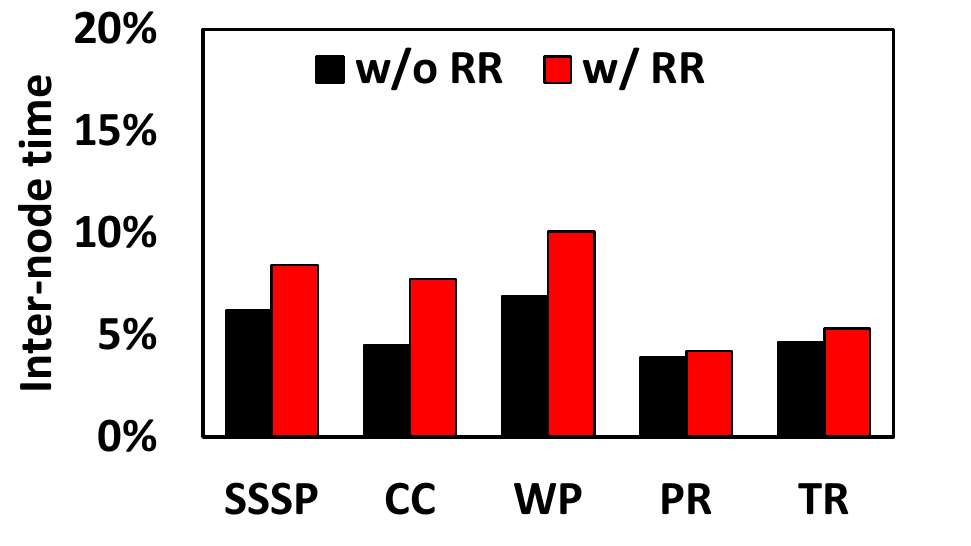}}
\end{minipage}%
\vspace{-0.1cm}
\caption{\textit{SLFE}'s RR effects on balance.}
\label{fig:imbalance}
\vspace{-0.5cm}
\end{figure}

\subsection{Discussion}
\label{sec:eva_discuss}

To verify the performance gain from optimizing the redundancies, we further study the number of computations during execution for SSSP, CC, and PR in Figure~\ref{fig:iterplot}. The reason for choosing these three applications is because of their representative converging trends among the five applications.

The SSSP initiates from a given root, and its number of computations dramatically increases as more vertices are reached along with the execution (Figure~\ref{fig:iterplot_sssp_fs} and~\ref{fig:iterplot_sssp_lj}). 
Redundant computations are reduced in the $pull$ mode. Hence, compared to the normal SSSP execution, \textit{SLFE}'s ramping-up curves reaches a much lower amount of computation. This phenomenon is caused by the ``start late'' approach, where intermediate updates are bypassed. Since our RR technique guarantees the correctness, both curves (w/RR and w/o RR) converge to the same point in the end. As aforementioned in Section~\ref{sec:rrrf}, push function activates all vertices to deliver ``unseen'' updates of inactive vertices in the pull$\rightarrow$push transition phase. 
We observe one such event (circled) in Figure~\ref{fig:iterplot_sssp_fs}), which only incurs a small amount of immediate computations to guarantee applications' functionality.
Figure~\ref{fig:iterplot_cc_fs} and~\ref{fig:iterplot_cc_lj} show that CC's number of computations is reducing along the converging. Similar to SSSP, CC's curves are finally merged in the end. 
In contrast, PR~\cite{graphchi,powergraph,powerlyra,geminigraph} keeps updating each vertex in the execution. 
As more EC vertices are detected in the execution, the ``finish early'' principle on these vertices dramatically reduces the total amount of computations (Figure~\ref{fig:iterplot_pr_fs} and~\ref{fig:iterplot_pr_lj}). In the end, \textit{SLFE} provides the same ranking result as compared to the one with the original algorithm.

In the end, we analyze \textit{SLFE}'s intra/inter-node balance of the five applications with all real-world graphs running on 8 nodes. Figure~\ref{fig:intranode} shows the intra-node case, where the baseline is the runtime achieved without stealing support. PR and TR are arithmetic-oriented algorithms that have a higher impact caused by intra-node work imbalance (work stealing reduces an average of 21\% runtime). In contrast, the work stealing scheme improves the performance of $min$/$max$-based applications by 15\%. As Figure~\ref{fig:iterplot} shows, the reduction is more effective in removing redundant computations on EC vertices. 

In the inter-node case, work balance is mainly ensured by the quality of the initial graph partitioning. 
Figure~\ref{fig:internode} shows, without RR, the average time difference between the earliest and latest finished nodes is less than 7\% across all applications, given the chunking-based partitioning approach~\cite{geminigraph}. 
With RR, $min$/$max$-based algorithms have a slightly higher inter-node imbalance compared to others. This is due to the imbalanced message passing in the push mode after redundancy optimization. In contrast, PR and TR have much less on-the-fly communication messages, because they always execute in the pull mode. Overall, \textit{SLFE}'s RR only increases an average of 2\% inter-node imbalance for all applications.

\section{Limitations}
\label{sec:limit}
\textit{SLFE} has two limitations. First, redundancy reduction guidance needs to be generated in the preprocessing phase. Even though generating this re-usable topological information for RR incurs extremely low cost, it is considered as overhead atop the original graph processing flow. Our future work is to further minimize the preprocessing overhead. Second, although not observed in our experiments, \textit{SLFE}'s efficient redundancy reduction could potentially incur workload imbalance across computation units when the amount of eliminated redundancies varies. For the intra-node case, we use work stealing to address this issue. However, it is challenging to address the potential inter-node load imbalance due to costly communication via network. In the future, we will investigate various inter-node work balancing schemes~\cite{mizan,Yanwww15} and integrate them into \textit{SLFE}.

\section{Related Work}
\label{sec:rw}
There are many distributed graph processing systems. Pregel~\cite{pregel} is the first one that proposes a vertex-centric programing model, which has been widely adopted by other graph systems~\cite{graphlab,graphchi,powergraph,powerlyra,powerswitch,geminigraph,gps}. Some existing work~\cite{xstream,chaos} developed the edge-centric graph processing engine that can sequentialize memory and I/O accesses. Other than the computation model designs, Powerlyra~\cite{powerlyra} and PowerSwitch~\cite{powerswitch} leveraged hybrid partitioning schemes and hybrid processing engines (sync/async) to accelerate graph analytics. However, none of these existing distributed frameworks aims to improve graph processing performance by optimizing the redundant computations. Unlike these approaches, \textit{SLFE} is the first one to optimize distributed graph processing with a novel redundancy-reduction design.  

Other than the distributed solution, GraphChi~\cite{graphchi} is a leading graph engine in a single PC; its parallel sharding window technique efficiently utilizes the secondary storage. Based on this scheme, Vora et al.~\cite{rajiv1} optimized GraphChi to only load edges with ``useful'' values. As claimed by the authors, this optimization relies on the particular re-sharding technique of disk-based systems, which is not applicable to systems with entire graphs stored in distributed memories. Kusum et al.~\cite{rajivhpdc} proposed a graph reduction method to improve computational efficiency of Galois~\cite{galois}. Such method performs iterative graph algorithms in a two-phase manner, which incurs an extra round of graph partitioning. Moreover, it cannot be applied to the distributed systems, because the preprocessing is the most expensive process in the distributed systems~\cite{vermavldb17,stantonkdd12,luvldb14,geminigraph}. By contrast, \textit{SLFE}'s solution is suitable for the distributed platforms, as it does not rely on any specific partitioning strategies, and does not need any extra partitioning effort.

\section{Conclusions}
\label{sec:conclusion}
In this paper, we propose \textit{SLFE}, a novel topology-guided distributed graph processing system. With the design principle of ``start late or finish early", \textit{SLFE} reduces redundant computations to achieve a higher performance. 
With the \textit{SLFE}'s system APIs, graph applications can easily harvest the performance benefits of RR-aware computation models. Experimental results show that \textit{SLFE} significantly outperforms three state-of-the-art distributed graph processing systems, delivering up to 74.8$\times$ speedup. More importantly, \textit{SLFE}'s redundancy detection and reduction schemes can be widely adopted by other systems.

{\footnotesize
\bibliographystyle{acm}
\bibliography{references}}

\end{document}